\documentstyle[12pt,aaspp4,psfig]{article}

\def\kms{km~s$^{-1}$}
\def\etal{{\it et al.} }
\def\msun{M_{\sun}}

\begin{document}
\hskip 2.5 truein To Appear: {\it Astronomical Journal}, March 1998 
\vskip 20pt
\title{The Complex Kinematics of the Neutral Hydrogen \\Associated with I~Zw~18}
\author{Liese van Zee\footnote{Jansky Fellow}}
\affil{National Radio Astronomy Observatory,\footnote{The National Radio 
Astronomy Observatory is a facility of the National Science Foundation,
operated under a cooperative agreement by Associated Universities Inc.} }
\affil{ PO Box O, Socorro, NM 87801}
\affil{lvanzee@nrao.edu}
\authoremail{lvanzee@nrao.edu}
\author{David Westpfahl}
\affil{Physics Department, NMIMT, Socorro, NM 87801}
\affil{dwestpfa@nrao.edu}
\author{Martha P. Haynes}
\affil{Center for Radiophysics and Space Research}
\affil{and National Astronomy and Ionosphere Center\footnote{The National
Astronomy and Ionosphere Center is operated by Cornell
University under a cooperative agreement with the National Science Foundation.}}
\affil{ Cornell University, Ithaca, NY 14853}
\affil{haynes@astrosun.tn.cornell.edu}
\author{John J. Salzer}
\affil{Astronomy Department, Wesleyan University}
\affil{Middletown, CT 06459--0123}
\affil{slaz@parcha.astro.wesleyan.edu}
\vfill
\eject
\begin{abstract}
We present the results of high velocity (1.3 \kms~channels) and high
spatial ($\sim$5\arcsec, or $\sim$ 250 pc at the distance of I~Zw~18) resolution
HI synthesis observations of the blue compact dwarf galaxy I~Zw~18 
to investigate the link between its unique evolutionary history and the
neutral gas distribution and kinematics.  The HI distribution is
extensive, with diffuse neutral gas extending to the northwest and south
of the main component.  This diffuse gas may be a remnant of the nascent HI cloud.
The kinematics of the I~Zw~18 system are complex, with 4 components identified:
HI--A, HI--C, HI--I, and HI--SX.  The gas associated with the main body, HI--A,
has a steep velocity gradient; although our analysis is hindered by poor spatial resolution
relative to the extent of the system, the main body appears to be undergoing solid
body rotation.  The optical condensation to the northwest of I~Zw~18 is embedded
in the common HI envelope, and is found to be kinematically separate from
the main body at a velocity of 740 \kms~(HI--C).  The interbody gas, HI--I,
connects HI--A and HI--C.  Finally, a large diffuse, kinematically distinct,
gas component extends at least 1\arcmin~to the south
of the main body (HI--SX), with no known optical counterpart. 
The peak of the gas column density coincides with the SE HII region in the main body; two other
HI peaks are associated with the NW HII region and an HII region in
the optical condensation to the northwest. In many respects, the HI properties of the
main body of I~Zw~18 (HI--A) are not unusual for dwarf galaxies; 
the peak column density, gas dispersion, M$_H$/L$_B$, and M$_H$/M$_T$ are
remarkably similar to other low mass systems.  The neutral gas
associated with I~Zw~18 is best described as a fragmenting HI cloud in
the early stages of galaxy evolution.
 
The derived gas distribution and kinematics are placed in the context of
the known star formation history of I~Zw~18.  In particular,
the neutral gas velocity dispersion is critical for calculating the abundance
of the HST detected \ion{O}{1} cloud.  While significantly affected by beam 
smearing in the presence of a steep velocity gradient, the derived gas velocity dispersion
in the main body of I~Zw~18 is approximately 12--14 \kms. Based on the present 
analysis, the \ion{O}{1} cloud has an oxygen abundance $\gtrsim$1/60th of solar, 
indicating that both the neutral and ionized medium are well--mixed.   
\end{abstract}

\section{Introduction}
\label{sec:intro}

With a predominately young stellar population (Hunter \& Thronson
\markcite{HT95}1995) and low elemental abundances (e.g., Skillman \&
Kennicutt \markcite{SK93}1993), the nearby blue compact dwarf galaxy
I~Zw~18 may be undergoing its first, or at most second, star
formation episode (Kunth \etal \markcite{KMM95}1995).  
Because of its fortuitous proximity ($\sim 10$ Mpc, so that 1
arcsecond = 48 pc), I~Zw~18 is a prime target for studies of the
early stages of the conversion of gas into stars in a galactic--sized
object.  Further, this galaxy provides a unique environment in which 
to test models of elemental enrichment and subsequent dispersal into the
interstellar medium (e.g., Marconi \etal \markcite{MMT94}1994; Roy \& Kunth
\markcite{RK95}1995; Kunth \etal \markcite{KMM95}1995).  
Despite the importance of I~Zw~18, surprisingly little is
understood about its neutral gas dynamics and kinematics; this paper
presents new high sensitivity, high resolution HI observations
of I~Zw~18 in order to investigate the link between the neutral
gas kinematics and the elemental enrichment of this enigmatic galaxy.

To understand fully the elemental enrichment of I~Zw~18, 
it is necessary to consider its local environment, evolutionary history, and
gas kinematics.  Recent HST observations provide excellent information
on its local environment and evolutionary history.  Of particular
relevance is the speculation that the high surface brightness compact
system also has one or more faint companions, and that the unusual
morphology and starburst activity result either from the merger of its
components, or alternatively from tidal interactions between
them. Speculations about possible companions to I~Zw~18 occur as far
back as the original discovery paper (Zwicky \markcite{Z66}1966).
Located spatially near I~Zw~18 are several faint, diffuse objects,
many of which appear to be roughly aligned towards the northwest
(Davidson \etal \markcite{DKF89}1989).  These stellar condensations
have been interpreted as the remnants of previous episodes of star
formation (Dufour \& Hester \markcite{DH90}1990; Dufour \etal
\markcite{DEC96}1996b), but the recent HST imaging observations of
Dufour \etal \markcite{DGSS96}(1996a) reveal that most of these
systems appear to be background galaxies.  Only one, component ``C''
in the nomenclature of Davidson \etal \markcite{DKF89}(1989), resolves
into stars in the HST images.  Subsequent spectroscopic observations
suggest that this object, also known as ``Zwicky's flare'', is a
physically--associated companion to I~Zw~18 (Dufour \etal
\markcite{DEC96}1996b; Petrosian \etal \markcite{PBCKL97}1997; this
paper).
 
Investigations of the neutral gas distribution and kinematics in
I~Zw~18 have been hampered by the difficulties of observing this
faint (in HI) and compact system.
Its HI flux density is relatively low, and disentangling HI
components that might be associated with the complex features evident
in the optical observations requires both high spatial and 
spectral resolution.  A preliminary 
understanding of the HI distribution and kinematics of I~Zw~18 were obtained 
from both WSRT (Lequeux \& Viallefond \markcite{LV80}1980) and 
multiconfiguration VLA observations (Viallefond \etal \markcite{VLC87}1987; 
Skillman \etal \markcite{SPGD96}1996).  These observations indicated that the 
compact optical galaxy is surrounded by a more extensive cloud of neutral hydrogen. 
Furthermore, the HI surface density peaks are associated with peaks in the 
optical surface brightness, although the precision of the coincidence 
remained unclear (Viallefond \etal \markcite{VLC87}1987). In addition, 
drawing attention to the possibility that the interstellar medium of I~Zw~18 
may be diluted with primordial gas,  Viallefond \etal \markcite{VLC87}(1987) 
identified several companion HI clouds.  These observations provided intriguing 
first looks at the HI distribution of this complex object, but also clearly 
indicated that higher spectral and spatial resolution observations would be 
necessary to disentangle the kinematics.

In this paper, we present the results of new
multiconfiguration HI synthesis observations of I~Zw~18 with the 
Very Large Array\footnote{The Very Large Array
is a facility of the National Radio Astronomy Observatory.}
providing both high velocity (1.3 \kms~channels) and high
spatial ($\sim$5\arcsec, or $\sim$ 250 pc at the distance of I~Zw~18) resolution.  
These observations permit a detailed analysis of the neutral gas kinematics and 
dynamics of the entire I~Zw~18 system.  
In particular, we are now able to compare kinematic
features in the ionized gas, such as the ``superbubbles''
identified by Martin \markcite{M96}(1996),  with features in the
neutral gas component.   In addition, ordered rotation has been 
detected throughout the main body of I~Zw~18, which is used to derive 
the dynamical mass of the system.  Further, the HI associated with
the faint optical object 25\arcsec~to the northwest of 
 I~Zw~18 (component C) is found to be kinematically distinct from the main body. 
 Finally, the combination of 
high spectral and spatial resolution observations permits a solid determination of
the neutral gas velocity dispersion, which is critical for the determination
of the O/H abundance in the HST detected \ion{O}{1} cloud (Kunth \etal \markcite{KLSV94}1994;
Pettini \& Lipman \markcite{PL95}1995). 

This paper is organized as follows.  The data acquisition and preliminary
analysis are presented in Section \ref{sec:data}.  The morphology and kinematics of the HI 
distribution are discussed in Section \ref{sec:morph}.  Section
\ref{sec:sfhist} contains a brief discussion of the past star formation activity in I~Zw~18;
we also speculate about the future possibilities of star formation in this
gas--rich system.  Our conclusions are summarized in Section \ref{sec:conc}.

\section{Data Acquisition and Analysis}
\label{sec:data}
Multiconfiguration HI observations of I~Zw~18 were obtained with
the VLA. Complementary optical imaging and spectroscopy
were obtained with the KPNO 0.9m, the Hubble Space Telescope,
and the Palomar 5m.  In this section, the procedures for observation 
and preliminary data reduction are discussed.

\subsection{HI Imaging Observations}
\label{sec:hidata}
Multiconfiguration observations of I~Zw~18 were obtained with the VLA
during 1993--1995.  The observing sessions totaled 3, 8, and 21.5 hours
in the D, C, and B configurations, respectively;  the total on--source
integration times are listed in Table \ref{tab:vlaobs}. 
During all observing sessions, the correlator was used in
2AD mode with a total bandwidth of 0.78 MHz, centered at 755 \kms.  
The on--line Hanning smoothing
option was selected, producing final spectral data cubes of 127 channels,
each 1.3 \kms~wide.  Standard tasks in AIPS (Napier \etal \markcite{AIPS}1983)
were employed for calibration and preliminary data reduction.  Each set
of observations was calibrated separately, using 3C 48, 
3C 147, and 3C 286 as flux and bandpass calibrators, and B0831+557 as the
phase calibrator.  The continuum emission was removed prior to CLEANing
and transformation to the xy plane.

The data were transformed with several weighting schemes and combinations
of configurations.  At least one map was made for each separate configuration;
in addition, many maps were made from the combined data sets using a
robust weighting technique (hereafter referred to as the ``BCD data cubes'').
The maps were created and CLEANed simultaneously 
with the AIPS task IMAGR (see Briggs \markcite{B95}1995 for a description of 
the robust weighting technique employed by IMAGR).  The robustness parameter
used in IMAGR and the beam sizes for a selected sample of the maps are listed 
in Table \ref{tab:maps}.  To boost the 
signal--to--noise in some of the maps, the data were smoothed in velocity
space by three channels, resulting in velocity resolution of 3.9 \kms~and
a decrease in the rms noise level by a factor of $\sim$1.7; these maps will 
subsequently be referred to as ``binned'' data cubes.  
The majority of the maps, however, retained the high velocity resolution 
of the original data.  All the data cubes were analyzed with the GIPSY 
package (van der Hulst \etal \markcite{GIPSY}1992).  

A continuum image was created from the line--free channels of the spectral 
data cube.  Due to the small bandwidth of these observations, the resultant
map has low signal--to--noise.  Very little continuum emission was detected, 
although a weak radio source ($<$ 3 mJy) appears to be located at approximately
09$^h$30$^m$30.2$^s$, +55\arcdeg27\arcmin48\arcsec~(1950) (also noted by 
Lequeux \& Viallefond \markcite{LV80}1980).  
The remaining radio sources appear to be background objects.  None are coincident
with the extensive HI envelope described in Section \ref{sec:mom0}.

\subsection{Optical Observations}
\label{sec:opt}
\subsubsection{Optical Images}
\label{sec:reg}
The optical imaging observations discussed in this paper were obtained from
several different telescopes.  As part of a larger program to image dwarf
galaxies on the KPNO\footnote{Kitt Peak National Observatory
is operated by the Association of Universities for Research in Astronomy, Inc. (AURA)
under a cooperative agreement with the National
 Science Foundation} 0.9m, a 5 minute R--band exposure of the 
11.7\arcmin~$\times$~11.7\arcmin~FOV around I~Zw~18 was obtained in 1995 March.  
The details of this observing run are discussed in van Zee \etal \markcite{vHS97}(1997b); 
the seeing was approximately 1.6\arcsec.  In addition,  
WFPC2 images of the I~Zw~18 system were extracted 
from the HST archive\footnote{Observations
with the NASA/ESA Hubble Space Telescope were obtained from the data archive at the 
Space Telescope Science Institute (STScI). STScI is operated by AURA under the NASA 
contract NAS 5--26555.} 
(PIs: Hunter GO--5309 and Dufour GO--5434). The observations were obtained on 1994  
October 29--31 and November 2--3.  The details of the HST observations are described 
in Hunter \& Thronson \markcite{HT95}(1995) and Dufour \etal \markcite{DGSS96}(1996a). 
The HST images were processed in the standard manner.  Images with the F555W 
filter were obtained by both observers, with different rotation angles and
placement of I~Zw~18 on the CCDs.  A mosaic of the two images was
created to extend the FOV around I~Zw~18.

Proper registration of optical images is crucial for comparison of the
neutral gas distribution with stellar and ionized gas features.  In the past, however,
it has been quite difficult to obtain accurate alignments.  For instance, 
in Dufour \& Hester (1990), the registration is based on a single star.
Unfortunately, their epoch 1950 coordinates are incorrect, by 5\arcsec~E and 2\arcsec~N;
when precessed to B1950, the original coordinates (epoch 1988.0) given in Davidson \etal 
\markcite{DKF89}(1989) agree with those listed in the APM catalog (Maddox \etal \markcite{MSEL90}1990).  
Even if their coordinates had been correct, it is impossible to derive a full astrometric 
plate solution with only one star.  With the advent of new star catalogues such as 
the APM catalog and the HST guide stars, however, it is now possible to easily obtain 
astrometric accuracy to better than 1\arcsec~for ground based images.  

Further difficulties are presented for the registration of HST images 
since few, if any, of the catalogued stars lie within the small FOV of the WFPC2 cameras.
While plate solutions are known for WFPC2 CCDs, we found systematic
differences of 2--3\arcsec~in the derived coordinates for objects in common between 
the 1994 October and November observations. We later discovered that the
header coordinates were incorrect for these images due to an error in one of the 
CCD reference positions (Fruchter, private communication).  This error was corrected 
shortly after the October and November 1994 observations, so additional WFPC2 observations 
of I~Zw~18 on 1995 March 1 were not affected.  This header problem is the
source of the offset between the HI and optical distributions seen in Skillman \etal 
\markcite{SPGD96}(1996) (Dohm--Palmer, private communication).  Due to the
apparent uncertainty in the WFPC2 astrometry, we elected to determine independently the
plate solutions for the WFPC2 images, via a bootstrap method.

The plate solution for the KPNO 0.9m image was determined based on 
the coordinates of 9 bright stars in the APM catalog (Maddox \etal \markcite{MSEL90}1990).
The coordinates of 8 faint objects located spatially near I~Zw~18 were then determined
from the KPNO image.  The majority of these objects are not point sources, but are rather 
faint background galaxies.  However, they were the only objects (other than a saturated 
star and the I~Zw~18 system) that were visible in both the KPNO and HST images.  The 
plate solutions for the HST images were then computed using the 8 faint objects.  
The derived plate solutions provide positional accuracy to better than 0.5\arcsec. 

\subsubsection{Optical Spectroscopy}
As part of a larger program to determine elemental abundances in
dwarf galaxies, optical spectra of the HII region in component C
 and the NW HII region of I~Zw~18 were obtained with
the Double Spectrograph on the 5m Palomar\footnote{Observations at the 
Palomar Observatory were made as part of a continuing cooperative agreement 
between Cornell University and the California Institute of Technology.} 
telescope on 1996 March 21.  A dichroic filter with a transition wavelength
of 5500 \AA~was used to split 
the light to the two sides (blue and red), providing complete spectral 
coverage from 3500--7600 \AA.   The blue spectra were acquired 
with the 300 lines/mm diffraction grating (blazed at 3990 \AA);  
the red spectra were acquired with the 316 lines/mm diffraction 
grating (blazed at 7500 \AA).  The slit was 120\arcsec $\times$ 2\arcsec.
A single 1800s observation at a position angle of 129\arcdeg~was obtained
(here, and throughout this paper, position angles refer to east of north).   
Further details of the observing run and data reduction are discussed
in van Zee \etal \markcite{vHS97}(1997b).

\section{HI Nomenclature, Distribution, and Kinematics}
\label{sec:morph}
In this section we discuss the HI distribution and kinematics of the I~Zw~18 system. In
the following discussion, the ``I Zw 18 system'' refers to the total HI distribution.
To avoid confusion, we will adopt the nomenclature of components as identified by
Davidson \etal \markcite{DKF89}(1989) throughout, with the modification that objects 
detected in HI will be preceded by ``HI''.  Thus, the rotating gas associated 
with the main body of I~Zw~18 will be designated HI--A.  The gas spatially associated
with component C will be designated as HI--C; the gas between
the two components will be designated the ``interbody gas'', or HI--I for short. 
The extended diffuse gas found primarily to the south and east and not associated 
with the preceding objects will be designated HI--SX.  Finally, Davidson \etal
\markcite{DKF89}(1989) note that the optical component A has two condensations;
we designate the associated HI structures as HI--A(SE) and HI--A(NW).

\subsection{Total HI Flux Density of the I~Zw~18 System}
\label{sec:flux}
To determine if the total HI flux density was recovered in 
the VLA observations, the lowest resolution (robustness = 5) data cubes of
each configuration and the combined configurations were examined.  For each
cube, the total flux density of the I~Zw~18 system was measured 
within a rectangular aperture of size 2.4\arcmin~$\times$ 3.0\arcmin, offset by
20\arcsec~S from the main body of I~Zw~18.   The derived global HI profile from
the combined configurations is remarkably similar to a profile obtained 
with the Green Bank 43 m (Haynes \etal \markcite{HHMRv98}1998), with 
deviations only on the order of the noise levels (Figure \ref{fig:gbflux}a).
The total flux density (2.97 Jy \kms) is also recovered within each configuration; the total
flux density of the combined BCD data cube was determined following the algorithm
of Haynes \etal \markcite{HHMRv98}(1998).  The ratio 
between the flux density measured in each separate configuration and the total flux density
of the combined BCD data cube is shown in Figure \ref{fig:gbflux}b.  
Even in the B configuration (with a minimum 
spacing of 1 k$\lambda$), the total flux density is recovered.

\subsection{HI Distribution}
\label{sec:mom0}
Selected channel maps of the lowest resolution BCD data cube are presented in
Figure \ref{fig:chans}.  As seen in the channel maps, the neutral gas associated 
with I~Zw~18 forms a complex system.  One main component is traced throughout 
the full data cube; we have designated this HI component as HI--A.  At low 
velocities (between 740 and 770 \kms), extended, diffuse structures are seen in 
the northwest.  This diffuse gas is spatially associated with the optical 
component C and also with the region between the two optical systems, A and C.

To further investigate the neutral gas distribution, the zeroth moment
of the lowest resolution binned (averaged over 3 channels) BCD data cube was 
computed.  To minimize the noise in the resultant maps, it has become 
standard practice to blank  regions of the cube which do 
not have signal prior to creating moment maps.  This method may result in loss of 
low signal--to--noise features, however.  Further, this method tends to 
introduce artificially steep gradients in the HI column density near 
the edges of the maps because the low column density (low signal--to--noise)
emission is blanked.  To avoid these problems, extreme care was taken
during the blanking process to retain all of the flux.  The blanking process 
began with a cube created from the combined B, C, and D configurations, but 
tapered at 13 k$\lambda$, resulting in a 17.4\arcsec~$\times$ 16.9\arcsec~beam.
This tapered cube is essentially equivalent to a smoothed cube derived from
the low resolution BCD data cube; tapering was preferred over smoothing because it
did not introduce edge effects.  The tapered cube was clipped at +/--1$\sigma$
(retaining everything above 1$\sigma$ and below --1$\sigma$) and the regions of 
emission were interactively blanked.  The blanked tapered cube thus creates a mask of
where the signal should be; this mask is then applied to the untapered BCD cubes.  
To confirm that no flux was lost in this process, the global HI profile 
of the blanked cube was checked.   The moments were calculated using 
a windowing function (signal must appear in more than one consecutive channel) 
to remove spurious noise spikes.  The resultant zeroth moment map is shown in 
Figure \ref{fig:mom0}.  The HI envelope is quite extensive, with diffuse gas 
structures extending as much as 1\arcmin~from the main component.  In addition, the 
clumps of gas identified by Viallefond \etal \markcite{VLC87}(1987) are recovered in 
this map, but they are found within a large, low column density neutral gas envelope 
rather than in isolation.  Excluding HI--SX and the other small features, the HI size 
(at a column density of 10$^{20}$ atom cm$^{-2}$) is approximately 60\arcsec~$\times$ 
45\arcsec~(2.9 $\times$ 2.2 kpc), at a position angle of 147\arcdeg.

To determine the contribution of the HI components (HI--A, HI--C, HI--I, and HI--SX)
to the total flux density of the I~Zw~18 system, the data cube was interactively clipped,
isolating each component.  The main HI component, HI--A, forms an elongated 
structure with two peaks in column density.  The flux emitted from this
rotating disk (see Section \ref{sec:rot}) was identified
based on its kinematic separation from the other components.  In the south, it 
is clearly distinct from HI--SX (see Section \ref{sec:piat}); in the north, it is
difficult to disentangle HI--A and HI--I, but extreme care was taken to
maintain spatial and velocity continuity.  The  total flux emitted from HI--A
is shown in Figure \ref{fig:flux}.  The HI profile of HI--A is remarkably symmetric;
the asymmetry in the global HI profile is thus due to the additional HI emission
to the northwest and south of HI--A.  Furthermore, separation of HI--A from the
larger diffuse gas envelope provides a very different picture of the gas density
distribution in this object than that illustrated in Figure \ref{fig:mom0}.  
The two main column density peaks remain, but the HI column density contours
 are relatively symmetric.  In
particular, there is no evidence of the precipitous contour crowding seen in 
the southern edge of HI--A as illustrated in Figure \ref{fig:mom0}.  This contour
crowding must be due to the overlap of HI--A with HI--SX in the moment map.

	In the northwest, HI--C appears as a secondary peak in the column density 
distribution (Figure \ref{fig:mom0}).  The ``interbody gas'', HI--I, connects
HI--C to HI--A.  Since it is impossible to accurately disentangle HI--I
and HI--C with the present spatial resolution, the combined flux density  of HI--I + HI--C
is shown in Figure \ref{fig:flux}.  The gas associated with HI--I and HI--C
contributes approximately 20\% of the total flux density (Figure \ref{fig:flux}a) and is 
localized in velocity space (Figure \ref{fig:flux}b).

  The low column density southern extension, HI--SX, is not easily seen in the 
individual channels (note that since the emission is at or below the 3$\sigma$ 
contour cut--off, the channel maps presented in Figure \ref{fig:chans}
do not include this region; spectral profiles are presented in Section \ref{sec:piat}). 
Nonetheless, HI--SX contributes $\sim$35\% of the total HI flux density in 
this system (Figure \ref{fig:flux}a); the flux density of this region was computed by 
summing the flux density remaining in the clipped data cube after excluding the HI--A, 
HI--C, and HI--I components.  Spatially, this includes all of the diffuse gas to 
the south and east of HI--A.  HI--SX is also localized in velocity space and accounts 
for some of the asymmetry of the global profile between 720 -- 760 \kms~(Figure \ref{fig:flux}b).

\subsection{Optical and HI Morphology}
\label{sec:hiopt}
A comparison of the HI and optical distributions is shown in Figure \ref{fig:opthi}. 
Due to the registration problems discussed in Section \ref{sec:reg}, previous
results suggesting an offset between the peaks in the HI gas distribution
and optical system (Dufour \& Hester \markcite{DH90}1990; Skillman \etal 
\markcite{SPGD96}1996) are not confirmed.
Instead, the HI column density maxima coincide with both optical components A and C.  
Higher spatial resolution data cubes reveal that the peak 
column density in the main body of I~Zw~18 is coincident with the HII region
known as A(SE) (Figure \ref{fig:hahi}). Further, a second peak in the
column density is roughly 
associated with the HII region A(NW) (the location of the peak optical surface brightness);
this secondary peak is slightly to the northwest of A(NW).
In addition, the peak column density in HI--C is associated with an HII region in the optical
component C.  Thus, as seen in other galaxies (e.g., van Zee \etal \markcite{vHSB97}1997a), 
local peaks of the neutral gas surface density in the I~Zw~18 system
are associated with the regions of active star formation. 

A main difference between this newer data set and the previous observations
of Viallefond \etal \markcite{VLC87}(1987) is the much larger extent of the HI
distribution. Thus, Figure \ref{fig:opthi} conveys a quite different picture
than Figure 1b of Dufour \& Hester \markcite{DH90}(1990). Contrary to the
previous impression, the ionized gas is {\it not} more extensive than the
neutral gas. Rather, the entire optical system is embedded in a diffuse,
irregular and clumpy neutral cloud, such as is commonly seen in 
normal dI galaxies.

\subsection{Velocity Field}
\label{sec:rot}
The velocity field (Moment 1) was calculated using a blanked data cube 
(see Section \ref{sec:mom0}) with moderate spatial resolution and the
full velocity resolution (Figure \ref{fig:vel}).  The derived neutral gas
velocity field is remarkably similar to that of the ionized gas (Petrosian 
\etal \markcite{PBCKL97}1997).  The velocity field is complex, but in the
region of HI--A a clear gradient is seen.  As also illustrated in the
position--velocity diagram (Figure \ref{fig:pv}), HI--C is seen as 
kinematically distinct from HI--A.    
 
In the interbody region, HI--I, between
HI--A and HI--C, the velocity field changes 
direction and has a much less steep gradient.  Though limited by 
our poorer spatial resolution, this result may reflect the same phenomenon 
as seen by Dufour \etal \markcite{DEC96}(1996b).  Inspection
of the individual channel maps in this region (Figure \ref{fig:chans}) 
suggests the presence of complex kinematics.  Whether HI--I
forms a bridge between HI--A and HI--C, or is the
result of gas bubbles or ``blow--out'' cannot be determined with
the present data set. It should also be noted that kinks and twists in the
velocity field are quite commonly seen in dwarf galaxies. The velocity field
displayed in Figure \ref{fig:vel} is not unlike that seen in the faint
dI galaxies mapped by Lo \etal \markcite{LSY93}(1993). 
 
In order to model the dynamics of HI--A, the contributions of HI--C, HI--I, and HI--SX
were clipped from the data cube (see Section \ref{sec:mom0}).  The resulting 
blanked cube contained only the main rotating component.  Because the derived
velocity field is only four beams across, it is impossible to constrain a standard 
tilted--ring model based on the kinematics alone.  Rather, a rough estimate of
the rotation curve was derived based on parameters obtained
from HI isophotal fits.  While the kinematic parameters are not
necessarily the same as the gas distribution parameters, 
such isophotal fits provide an estimate of the center coordinates, position angle,
and inclination angle of the rotating gas disk.
An ellipse centered at 09$^h$30$^m$30.0$^s$, 
+55\arcdeg27\arcmin49.0\arcsec, with a position angle of 147\arcdeg~and 
an inclination angle of 55\arcdeg~best fit the HI mass distribution.
The center of this ellipse is coincident with the NW HII region.
The shape of the outermost ring used in the tilted-ring models is illustrated 
in Figure \ref{fig:vel}.  The inclination angle was derived by assuming a
planar HI distribution; if the HI distribution is thick, as is seen in other 
dwarf galaxies, the inclination will be smaller than estimated here.  In addition,
a systemic velocity of 762 \kms~(the center of the HI--A profile)
was adopted.  The derived rotation curves for both
sides are shown superposed with a position--velocity diagram (from an
unclipped cube) in Figure \ref{fig:pv}.  Due to the poor spatial resolution
of these observations relative to the size of HI--A, it is still unclear
whether the steep velocity gradient is intrinsic to HI--A, or is caused
by beam smearing.  However, the steep, solid body--like, 
velocity gradient across component HI--A is suggestive.  Component
HI--C appears as a kinematically distinct object offset 25\arcsec~to the NW
from the kinematic center of HI--A, and at a velocity of 740 \kms, similar
to that found by Petrosian \etal \markcite{PBCKL97}(1997).

A slight change in the velocity gradient may occur across HI--A. 
In the kinematic models, the receding side has a velocity gradient 
of 6.0 \kms~arcsec$^{-1}$ while the 
gradient in the approaching side is much shallower, only 2.2 \kms~arcsec$^{-1}$.
Similarly steep and changing velocity gradients have been seen in the 
ionized gas (e.g., Skillman \& Kennicutt \markcite{SK93}1993;
Martin \markcite{M96}1996; Petrosian \etal \markcite{PBCKL97}1997).
Skillman \& Kennicutt \markcite{SK93}(1993) hypothesized that
the changing velocity gradient might indicate that the main body
of I~Zw~18 is actually two colliding clumps of gas.
The present data set does not have sufficient spatial resolution
to resolve this issue.   

Despite the above caveats, a dynamical mass, based
on the assumption that the main body of I~Zw~18 is a gravitationally bound system, 
was computed from the maximum velocity of the derived rotation 
curve, $\sim$44 \kms~at a radius of 12\arcsec.   The dynamical mass
of the main body thus computed
is approximately 2.6 $\times$ 10$^8$ M$_{\odot}$.
Similar to other dwarf galaxies (e.g., Skillman \etal \markcite{SBMW87}1987; 
van Zee \etal \markcite{vHSB97}1997a), the ratio of the HI to dynamical
mass in the main body of I~Zw~18 is 0.1. 

\subsection{Velocity Dispersion}
\label{sec:sig}
The spectral profiles across HI--A, HI--I, and HI--C are 
shown in Figure \ref{fig:prfs}.  The majority of these profiles
are asymmetric, and suggestive of two Gaussian components.
Based on the (unpublished) Viallefond \etal VLA maps, Kunth \etal \markcite{KLSV94}(1994)
quote a velocity dispersion of $\sim$ 19 \kms~in the region of a GHRS
absorption spectrum [centered on 09$^h$30$^m$30.2$^s$, +55\arcdeg27\arcmin48\arcsec (1950),
the NW HII region; note that the published coordinates differ 
from those listed in the HST archives]. 
Based on either a second moment map or single Gaussian component fits to the
spectra, the present data result in a similar velocity dispersion at this location.
The highly asymmetric profiles, however, suggest that these approaches are too 
simplistic.  Asymmetric profiles are seen across most of HI--A 
(the 3 across by 6 up rectangle in the lower left of Figure \ref{fig:prfs}).  These profiles are
well fit by a double Gaussian model, where the second Gaussian is offset by $\sim$30 
\kms~from the first.  Given the poor spatial resolution of these observations relative
to the separation of the HI--A(NW) and HI--A(SE) column density peaks, these asymmetric 
profiles may be the  result of beam smearing.  Beam smearing can introduce an overlap of 
the two gas clumps in the spectra in an asymmetric manner in the presence of
a steep velocity gradient.  In the position--velocity 
diagram (Figure \ref{fig:pv}), the combination of a steep rotational gradient and 
moderate gas dispersion can be seen to introduce such asymmetric features.

If the two Gaussians arise from the overlap of HI--A(NW) and HI--A(SE), the gas 
velocity dispersion at each position may be computed from the width of the main Gaussian.
Typical dispersions range between 12--14 \kms~for the main Gaussians (this agrees with the 
dispersion in regions where the profile is more symmetric, such as the lower left hand 
corner of Figure \ref{fig:prfs}).   The derived velocity dispersions are upper
limits on the gas velocity dispersion since even the decomposed profiles may have
significant rotational components given the steep velocity gradient in this
system.  Thus, the fact that the derived gas dispersions are slightly higher than those 
seen in other dwarf galaxies (e.g., Lo \etal \markcite{LSY93}1993; Young \& Lo \markcite{YL96}1996; 
van Zee \etal \markcite{vHSB97}1997a) should not be considered significant.

\subsection{Interpretation of Components}
\label{sec:piat}
HI gas is seen to spatially coincide with both optical components (A and C)
of the I~Zw~18 system.  It is reasonable to assume that this gas is associated
with the stellar systems in these regions.  The HST images provided the
first solid indications that components A and C are physically associated
 (Dufour \etal \markcite{DGSS96}1996a); this concept was  
further supported by optical redshift determinations (Dufour \etal \markcite{DEC96}1996b;
Petrosian \etal \markcite{PBCKL97}1997; 
this paper).  The present HI data suggest that both components reside within
an extensive HI envelope.  While HI--C appears to be kinematically distinct 
from HI--A, the interbody gas, HI--I, between HI--A and HI--C has a smoothly 
changing velocity field which connects the two systems.

Other authors have used the term ``companion dI galaxy'' to describe optical component
C (e.g., Dufour \etal \markcite{DEC96}1996b) based on its similar redshift to component A.  We
have avoided such terminology in this paper to prevent the impression that these 
are tidally interacting systems.  The picture developed here, based on the neutral 
gas distribution and kinematics, is that of a fragmenting HI cloud with at least two sites of
star formation activity (A and C).  As both components appear to be very
young (see Section \ref{sec:stars}), it is likely that they are separate pieces
of a single, rapidly evolving system.

A large percentage ($\sim$35\%) of the HI gas in the I~Zw~18 system appears as
a diffuse envelope, unrelated to known optical counterparts.  This 
diffuse gas is localized in velocity space, but does not connect smoothly
to HI--A.  Spectral profiles of the gas in the southern extension (HI--SX)
are shown in Figure \ref{fig:sprfs}.  The gas in this region is centered
at a velocity of 739 \kms, with a velocity dispersion of approximately 10 \kms, 
similar to that found in HI--A.  However, as is seen in the center two panels
of the top row of Figure \ref{fig:sprfs}, HI--SX appears to be kinematically
distinct from HI--A (which appears at $\sim$800 \kms~in these panels).
The discontinuity in velocity field between
HI--SX and HI--A suggests that HI--SX is not a tidal tail, stripped from
HI--A.  Rather, HI--SX may be a remnant of the nascent HI envelope
which perhaps produced both HI--A and HI--C and, eventually, their optical
counterparts.  

\section{Star Formation History}
\label{sec:sfhist}
In this section we summarize  previous results on the evolutionary
history of the I~Zw~18 system in light of the new information of the neutral
gas distribution and kinematics.  A summary of the physical properties
of I~Zw~18 is presented in Table \ref{tab:props}.

\subsection{Ionized Gas}
The two bright knots of star formation in component A of the I~Zw~18 system
are separated by $\sim$5.6\arcsec~(Zwicky 1966).  High spatial resolution 
H$\alpha$ images of I~Zw~18 reveal that the NW HII region, A(NW), resolves 
into a small HII region and a shell--like structure that encompasses a 
stellar association (Hunter \& Thronson \markcite{HT95}1995; Figure \ref{fig:hahi}).
The SE HII region, A(SE), remains compact, even in the high spatial 
resolution HST images.  One faint HII region  has also been identified 
in component C.  Aside from the compact HII regions, large filaments and 
loops of ionized gas are seen to extend several hundreds of parsec from the 
center of component A.  

Two of these loops appear to be expanding ``superbubbles'' of
ionized gas, with an expansion rate of $\sim$ 30--60 \kms~(Martin \markcite{M96}1996).
The existence of such superbubbles suggests that I~Zw~18 may not be
undergoing its first episode of star formation; further, the superbubbles
provide a mechanism to remove the enriched material of previous generations
of star formation, perhaps explaining the low abundance nature of I~Zw~18.
The position--velocity diagram of the neutral gas in the vicinity of
one ``superbubble'' is shown in Figure \ref{fig:pv7}.  The slice is centered
on A(NW) and cut at a position angle of 7.7\arcdeg, equivalent to the
positioning of Martin's slit 1.  Two Doppler ellipses were identified in
the ionized gas, one at a spatial offset of 10--20\arcsec~to the SW of A(NW),
and one 10--15\arcsec~to the NE.  Any neutral gas associated 
with the more prominent feature in the ionized gas to the SW is not 
detectable in the present data set because the neutral
gas column density drops precipitously in this region.  Furthermore, the
peak flux of ionized gas occurs at $\sim$820 \kms, which is at the edge 
of the HI bandpass.  However, a hint of the NE kinematic feature is 
visible in the neutral gas position--velocity diagram (Figure \ref{fig:pv7}),
at very low signal--to--noise. 

\subsection{Stellar Component}
\label{sec:stars}

Spectra of the HII regions are dominated by emission lines, with only  
weak, blue continuum emission (e.g., Skillman \& Kennicutt \markcite{SK93}1993).
The very blue stellar continuum was the first indication that the main 
body might consist of young, massive stars (Sargent \& Searle \markcite{SS70}1970).  
Since the emission lines are strong relative to the continuum 
emission in this system, optical broad band images are also dominated by 
the emission lines.  Recently, emission--line corrected
broad band magnitudes have been obtained: U = 15.727 $\pm$ 0.053,
B = 16.608 $\pm$ 0.027, V = 16.637 $\pm$ 0.025 (Salzer \etal 
\markcite{SESS98}1998).  Based on the Bruzual \& Charlot 
\markcite{BC93}(1993) stellar evolution models, the derived colors 
are consistent with a very young burst of star formation, $\leq$10 Myr old. 

With the advent of HST, detailed studies of the stellar composition of 
components A and C are now possible.  In their analysis of the
stellar population of component A, Hunter \& Thronson \markcite{HT95}(1995)
subdivide the main body of I~Zw~18 into three regions: the shell stars within
the shell of H$\alpha$ emission in the NW, the stars associated with
the SE HII region, and the remaining stars associated with
I~Zw~18.  As might be expected, young ages are determined for the
regions associated with the two HII regions.  The brightest stars in the 
galaxy are located within the H$\alpha$ shell in the NW.   The 
absence of stars older than 6 Myr in this region suggests that the shell 
stars are young, with typical ages of 2--5 Myr (Hunter \& Thronson \markcite{HT95}1995).  
Further, the lack of bright and red stars suggests that the southern HII region may be even
younger than 2 Myr; that is, the majority of the 
stars in the south still appear to be on the main sequence (Hunter \& 
Thronson \markcite{HT95}1995).  In contrast to these two HII regions, the 
general field stars (those not confined to either the NW or SE HII regions) 
appear to be slightly older, and perhaps are not coeval (Hunter \& Thronson 
\markcite{HT95}1995).  Based on an independent data set, Dufour \etal 
\markcite{DGSS96}(1996a)  found several outlying blue B-- and A--supergiants, 
indicating that star formation began at least 30 Myr ago, and perhaps as much 
as 50 Myr ago, in the main body of I~Zw~18.

As mentioned previously, HST observations of the optical component C
(Zwicky's flare) by Dufour \etal \markcite{DGSS96}(1996a) resolve the
object into stars,
allowing the construction of a color--magnitude diagram.
The presence of a well defined upper main sequence indicates that the 
blue stars have an approximate age of 40 Myr. However,
numerous red stars are also seen.  These red stars correspond
to 100--300 Myr isochrones, suggesting that component C has been
forming stars for a significantly longer time period than component A
(Dufour \etal \markcite{DGSS96}1996a).  Further evidence 
suggesting that the stellar population in C has an
older age comes from the hint of Balmer absorption features in the
optical spectrum of its HII region.  
Figure \ref{fig:optspec}
shows the extracted spectra obtained with the 5m Palomar telescope
for both the HII region in C and the NW HII region in A, A(NW); 
the spectra confirm that both HII regions lie at the same redshift.
In addition, only H$\alpha$ and H$\beta$ are seen in the
emission spectrum of component C, suggesting a soft radiation field. 
The very weak absorption features at the wavelengths of H$\gamma$,
H$\delta$, and H$\epsilon$ also suggest that this HII region is
ionized by an early to intermediate  B star.  Because of the
absence of diagnostic lines, it is not possible to
constrain the elemental enrichment of component C.

\subsection{Elemental Enrichment}

The two bright knots of star formation in component A 
have the lowest oxygen abundance
of any known star forming galaxy, approximately 1/50th of solar (e.g.,
Skillman \& Kennicutt \markcite{SK93}1993).  Further, the derived 
metallicities are remarkably similar in these two
knots of star formation, suggesting a coupled and relatively limited star
formation history.  Alternatively, the products of stellar nucleosynthesis 
may not yet be mixed with the ionized medium, particularly if they are 
released in a hot, highly--ionized phase.  

Despite the extensive metallicity measurements of emission--line galaxies
(e.g., Masegosa \etal \markcite{MMC94}1994), not one has yet been found
to have a lower oxygen abundance than the HII regions in I~Zw~18.
Kunth \& Sargent (1986) argue that the absence of similarly
low metallicity objects is not a result of observational
selection, but due to the self--enrichment of the HII regions observed.
While such an argument is compelling, no such self--enrichment has
been observed in actively star forming dwarf galaxies.  In fact, multiple abundance 
measurements in dwarf galaxies almost universally result in negligible abundance
gradients (e.g., SMC and LMC, Pagel \etal \markcite{PEFW78}1978; NGC 4214,
Kobulnicky \& Skillman \markcite{KS96}1996; NGC 1569, Kobulnicky \& Skillman 
\markcite{KS97}1997), suggesting either that the mixing process is
very efficient in these small galaxies, or that prompt self--enrichment is
negligible.

The discovery of a low abundance \ion{O}{1} cloud around I~Zw~18 
(Kunth \etal \markcite{KLSV94}1994), however, suggests that the ionized
material may have been significantly enriched, perhaps by self--pollution.
Further, an apparently sharp abundance gradient between the ionized and neutral gas
led to speculation about {\it inefficient} mixing mechanisms
in dwarf galaxies (Roy \& Kunth \markcite{RK95}1995). 
The \ion{O}{1} line was saturated in the GHRS spectrum, however,
which meant that the oxygen abundance derivation for the \ion{O}{1} cloud required an assumption
about the neutral gas dispersion factor in this region (Kunth \etal \markcite{KLSV94}1994).
In the absence of an independent estimate of the velocity dispersion,
 Pettini \& Lipman  \markcite{PL95}(1995) illustrated that the \ion{O}{1} absorption spectrum
was consistent with abundances between solar and 1/1000 of solar.
 In the original derivation of the abundance, Kunth  \etal \markcite{KLSV94}(1994)
assumed a dispersion factor $b=\sqrt{2}\sigma$ of 27 \kms~based on the (unpublished) low
velocity resolution HI data set of Viallefond \etal  As seen in Section
\ref{sec:sig}, however, the steep velocity gradient across the main body
introduces significant broadening of the HI line profile.
With a derived neutral gas dispersion factor of 13--14 \kms, 
a $b$ of 18--20 \kms~appears to be appropriate for the region of the GHRS 
absorption spectrum. Combining the analysis of Pettini \& Lipman \markcite{PL95}(1995)
with this revised value for the dispersion factor allows us to derive an 
abundance for the O I cloud of $\gtrsim$1/60th of solar.
Higher spatial resolution HI observations will be necessary, however,
to confirm the derived velocity dispersion; it is quite possible that 
the asymmetric HI profiles arise from more complex gas kinematics than  
modelled in this paper.

This new analysis of the abundance of the \ion{O}{1} cloud
suggests that the abundance gradient, if it exists, is very
shallow between the ionized and neutral medium.  Given that
the color--magnitude diagrams indicate that star formation
has  been underway for the last 30--50 Myr in
I~Zw~18, products of stellar nucleosynthesis must have
been released into the interstellar medium as the massive
stars evolved. It is unclear how much enriched material
from past star formation episodes has been
ejected and mixed into the interstellar medium, but
the mixing and dispersal process appears to be quite efficient 
throughout both the ionized and neutral medium.
  
One significant question in such a low mass galaxy as I~Zw~18 is whether
the enriched materials are retained by the system.  The superbubbles
of Martin \markcite{M96}(1996) could be the result of 
energetic processes expelling ionized gas into the intergalactic
medium.  Whether that gas carries the enriched materials with
it, and whether that material may eventually fall back into the
galaxy, is unknown at this time.   

\subsection{Empirical Star Formation Thresholds}
\label{sec:thresh}

Comparisons between neutral gas distributions and sites of active star formation
suggest that the local gas column density plays a critical role in
regulating star formation activity in dwarf galaxies (e.g., Skillman \etal 
\markcite{SBMW87}1987; Taylor \etal \markcite{TBPS94}1994; van Zee \etal 
\markcite{vHSB97}1997a).  While both the atomic and molecular gas components should
be considered, the common tracer of
molecular gas,  CO, has proven to be extremely difficult to detect in 
metal--poor, low mass galaxies (e.g., Elmegreen \etal \markcite{EEM80}1980).  
Observations of the CO in I~Zw~18 with the IRAM 30m have resulted in an upper limit
of L$_{CO}$ $<$ 1.6 K \kms~kpc$^2$, when scaled to D=10 Mpc (Arnault \etal \markcite{ACCK88}1988).  Even if CO were detected,
the conversion factor from CO column density to the more abundant
H$_2$ column density is highly uncertain in metal--poor systems
(e.g., Verter \& Hodge \markcite{VH95}1995; Wilson \markcite{W95}1995).
In the absence of such measurements, we have elected to apply no
correction for the contribution of molecular gas in I~Zw~18.
Thus, while a correction for atomic helium (N$_{gas}$ = 4/3 N$_{HI}$) is
applied, the gas column densities quoted below are lower limits. 

The HI and H$\alpha$ distributions for I~Zw~18 are shown 
in Figure \ref{fig:hahi}.  The peak column density of 
3.02$\times$10$^{21}$ atoms cm$^{-2}$ is coincident 
with the SE HII region in the main body of I~Zw~18. Corrected for an inclination 
of 55\arcdeg~(see Section \ref{sec:rot}) and the neutral helium content, 
this corresponds to a peak neutral gas column density of 2.3$\times$10$^{21}$ atoms 
cm$^{-2}$ at a spatial resolution of $\sim$250 pc.  This peak column density is high, 
but not atypical of column densities seen in other star forming dwarf galaxies 
(e.g., Taylor \etal \markcite{TBPS94}1994; van Zee \etal \markcite{vHSB97}1997a).  
An observed column density of 1.8$\times$10$^{21}$ atoms cm$^{-2}$ (inclination corrected
neutral gas column of 1.4$\times$10$^{21}$ atoms cm$^{-2}$) well delineates 
the two main H$\alpha$ knots, suggesting that the empirical gas column
density threshold of 10$^{21}$ atoms cm$^{-2}$  (Skillman \markcite{S87}1987)
is surpassed throughout the active star forming regions.

The origin of the empirical critical threshold density seen in many
dwarf galaxies is the subject of much debate.  One possibility is
that this threshold may correspond to a critical column density of dust 
necessary to shield molecular gas from UV radiation (Federman \etal \markcite{FGK79}1979).  
If so, one might expect this threshold value to be a function of elemental abundance 
(Skillman \etal \markcite{SBMW87}1987).  The preliminary results on the peak 
column density in I~Zw~18 by Viallefond \etal \markcite{VLC87}(1987) suggested 
that a correlation might be found between high peak HI column density and 
low metallicity.  The new observations, however, indicate that the peak column density
in I~Zw~18 is not unusually high.  High spatial resolution observations of 
systems with a range of metallicities and UV fields will be necessary to 
fully investigate  this issue, however.

\section{Summary and Conclusions}
\label{sec:conc}

In summary, we have obtained new high sensitivity,
high spatial and spectral resolution HI synthesis
observations with the VLA that provide new insight into the neutral gas distribution 
and kinematics of I~Zw~18. Our results and conclusions are summarized as follows.

(1) The total HI line flux densities detected in the B, C, and D configurations are
consistent with that derived from the large beam 43~m Green Bank antenna
observations of Haynes \etal \markcite{HHMRv98}(1998), indicating that the 
total flux density (2.97 Jy--\kms) is recovered in the VLA observations.
The mass of HI present in the system amounts to $7.0 \times 10^7\msun$,
comparable to that seen in other nearby dwarfs (Lo \etal \markcite{LSY93}1993).
At least 4 kinematically distinct components are visible in the HI data cube.

(2) The HI size (at a column density of 10$^{20}$
atom cm$^{-2}$) is approximately 60\arcsec~$\times$ 45\arcsec 
~(2.9 $\times$ 2.2 kpc), 
at a position angle of 147\arcdeg. In its irregularity of shape and
clumpiness, the HI distribution is quite similar to those seen in
faint dI galaxies of comparable HI mass (Lo \etal \markcite{LSY93}1993).  

(3) The main HI component, HI--A, forms an elongated, asymmetric structure.
The peaks of the HI distribution are coincident with the regions
of highest optical surface brightness. The HI column density reaches
a maximum at 3.02 $\times$ 10$^{21}$ atoms cm$^{-2}$, coincident with the SE
HII region in the main body, A(SW). A peak of somewhat lower column
density is associated with the other bright HII region, A(NW). 
As in other dwarf systems, the neutral gas column density exceeds  
10$^{21}$ atoms cm$^{-2}$ throughout the active star forming regions.

(4) Although the overall kinematics of the HI associated with I~Zw~18 
are complex and the spatial resolution limited, evidence of
solid body rotation is detected across the component HI--A.
If the observed velocity field is truly reflective of solid
body rotation, the dynamical mass of 
HI--A is approximately 2.6 $\times$ 10$^8$ M$_{\odot}$.
As seen in many other dwarf galaxies, the HI--to--dynamical mass is
approximately 0.1, indicating a substantial contribution of dark matter.

(5) A third peak in the HI column density coincides with an HII 
region in the optical component C,
offset 25\arcsec~to the NW from the kinematic center of HI--A.
HI--C is kinematically distinct from HI--A at a velocity of 740 \kms.

(6) The observed spectral profiles in the region of HI--A
are strongly asymmetric, but well fit by two Gaussians separated by 
$\sim$30 \kms. The asymmetric profiles are probably a result of beam
smearing in the presence of a steep  velocity gradient.
After account is taken of the beam smearing, 
the derived gas dispersion is approximately 12--14 \kms, similar to what 
has been found in other dwarf galaxies, but significantly lower
than that estimated previously for I~Zw~18, based on poorer 
resolution data. Adoption of a lower dispersion for the region
of the GHRS spectrum gives a gas dispersion factor
$b$ of 18--20 \kms. Combining this velocity dispersion factor with
the analysis of Pettini \& Lipman \markcite{PL95} (1995),
the derived abundance of the
HST detected \ion{O}{1} cloud is $\gtrsim$1/60th of solar.

(7) While kinematically distinct, the HI associated with the optical
component C, HI--C, appears to be connected to HI--A by interbody gas with a 
smoothly changing velocity gradient.  This apparent bridge
could  be the result of gas bubbles or ``blow--out''.

(8) About 35\% of the total HI flux density arises in a diffuse extended component,
HI--SX, not coincident with any optical emission. 
HI--SX is also localized in velocity space and accounts 
for some of the asymmetry of the global profile between 720 -- 760 \kms.
In contrast to impressions conveyed by previous lower sensitivity 
observations, the optical system in I~Zw~18 is embedded in an extensive
cloud of low column density ($\sim$ 5 $\times$ 10$^{19}$ atoms cm$^{-2}$)
neutral gas. The neutral gas is therefore not embedded in a hot envelope,
but rather the opposite situation occurs. The current data set remains
limited in spatial resolution and signal--to--noise and does not allow
discrimination amongs models of the overall HI distribution as
a single nascent envelope or the result of a merger or tides. However,
no velocity gradient is evident across HI--SX. The lack of a gradient
might result from geometry if the cloud is a disk or tidal tail seen 
face--on. More likely however, HI--SX is neither face--on nor
disk--like. Similarly patchy distributions are seen on 
comparable linear scales in the faintest
dwarf galaxies mapped by Lo \etal \markcite{LSY93}(1993) which,
those authors argue, are dominated by turbulence, not rotation.

(9) A new optical spectrum of the brightest HII region associated
with the optical component C reveals the presence of only
H$\alpha$ and H$\beta$ in emission plus 
weak Balmer absorption features, implying that the HII region
is ionized by an early to intermediate B star. 
In accord with the HST color--magnitude diagram
reported by Dufour \etal \markcite{DGSS96}(1996a), the
spectral characteristics suggest that the stellar
component of C is older in age than the bursting population
seen in A.

In many respects, the HI distribution and kinematics of I~Zw~18 are
not unusual for dwarf galaxies.  For instance, when the other components
are excluded, properties such as peak column density, gas dispersion, 
M$_H$/L$_B$, and M$_H$/M$_T$ for component A, are remarkably similar to those
found in other low mass systems (e.g., Skillman \etal \markcite{SBMW87}1987; 
Lo \etal \markcite{LSY93}1993; van Zee \etal \markcite{vHSB97}1997a).  
The main exception is the existence of diffuse gas extending to the south 
of the main body.   ``I~Zw~18'' may actually consist 
of a conglomeration of low mass clumps of gas contained within 
a larger diffuse envelope.

Such a conglomeration might explain the unique properties of I~Zw~18,
such as the extremely low oxygen abundance and evidence for a very
recent onset of star formation in both the main body and component C.
If I~Zw~18 is in the process of accreting small gas clouds, such as
those originally identified by Viallefond \etal \markcite{VLC87}(1987),
the interstellar medium will be diluted by this pristine material, 
resulting in a low oxygen abundance.  Further, the very process of
accretion may have induced the present star formation activity, once
the local gas density exceeded the critical density.  While such a
scenario has simplistic appeal, the absence of a strong abundance gradient 
in the neutral and ionized medium suggests a coupled and unevolved
star formation history, rather than the slow accretion of pristine
gas.

To resolve fully the effect of the neutral gas distribution and kinematics
on the star formation history of I~Zw~18, higher spatial resolution
observations with comparable spectral resolution will be necessary. 
In particular, with the present spatial resolution it is impossible to
determine whether the steep and changing velocity gradient in HI--A 
is due to additional fragmentation of the gas cloud, or the result of
gas bubbles and ``blow--out.''  Further,  determination
of the gas velocity dispersion was significantly hampered by beam effects  
in the presence of a steep velocity gradient.  Given that the
total flux density  was recovered in the B configuration alone, such high spatial 
resolution observations are possible, and indeed necessary, to
understand the neutral gas kinematics in this unique galaxy.

\acknowledgements
We acknowledge Daniel Puche, who participated in the D and C configuration
observations and the early stages of this project.
We thank Lisa Young and Evan Skillman for many enlightening conversations 
about dwarf galaxies and gas kinematics. We acknowledge the financial 
support by NSF grants AST90--23450 and AST95-28860 to MPH, AST95--53020 to JJS and
from the New Mexico Space Grant Consortium to DW.

\begin{table}
\dummytable\label{tab:vlaobs}
\end{table}

\begin{table}
\dummytable\label{tab:maps}
\end{table}

\begin{table}
\dummytable\label{tab:props}
\end{table}

\vfill
\eject
\psfig{figure=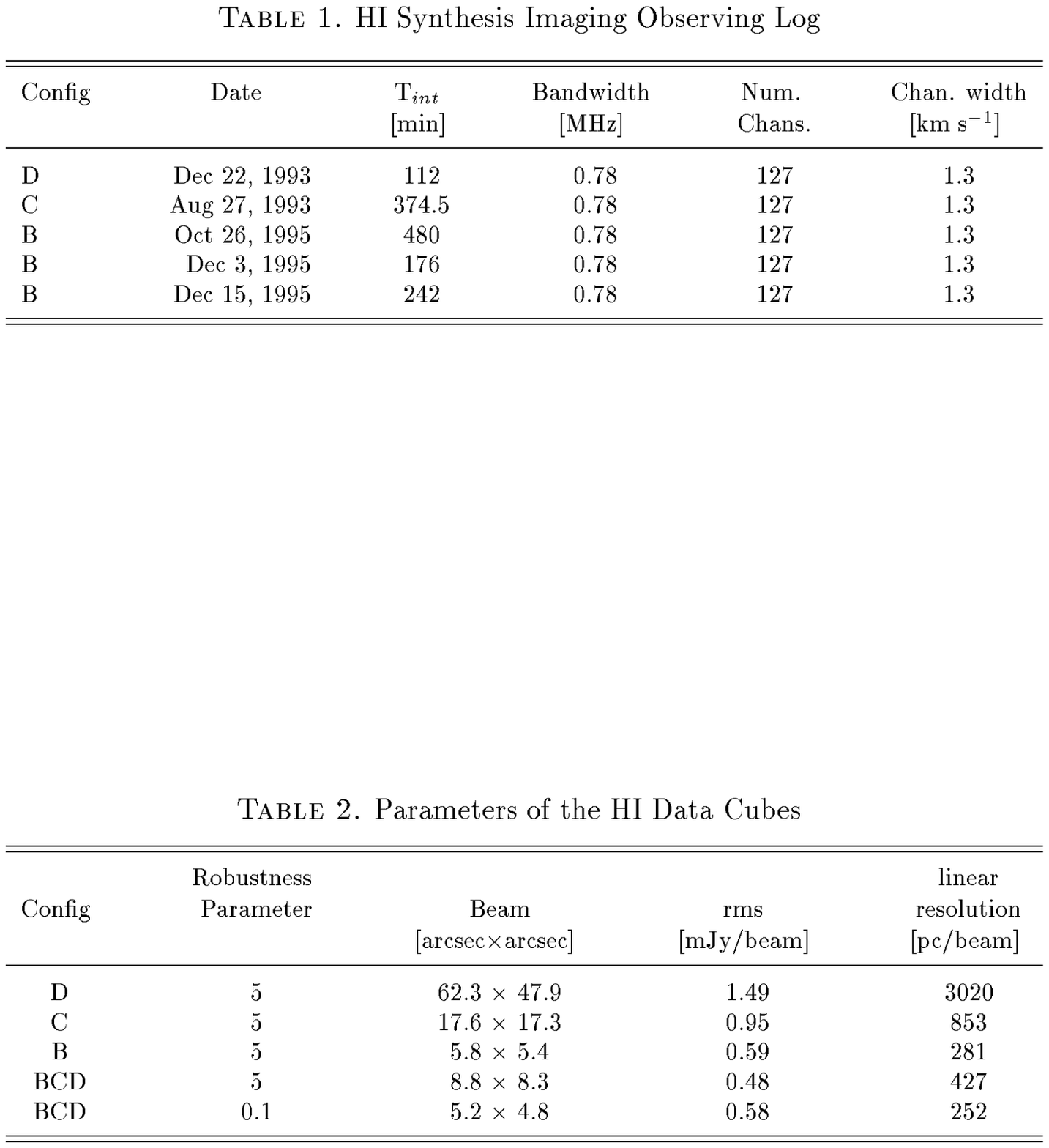,width=7.in,bbllx=50pt,bblly=150pt,bburx=600pt,bbury=700pt,clip=t}

\psfig{figure=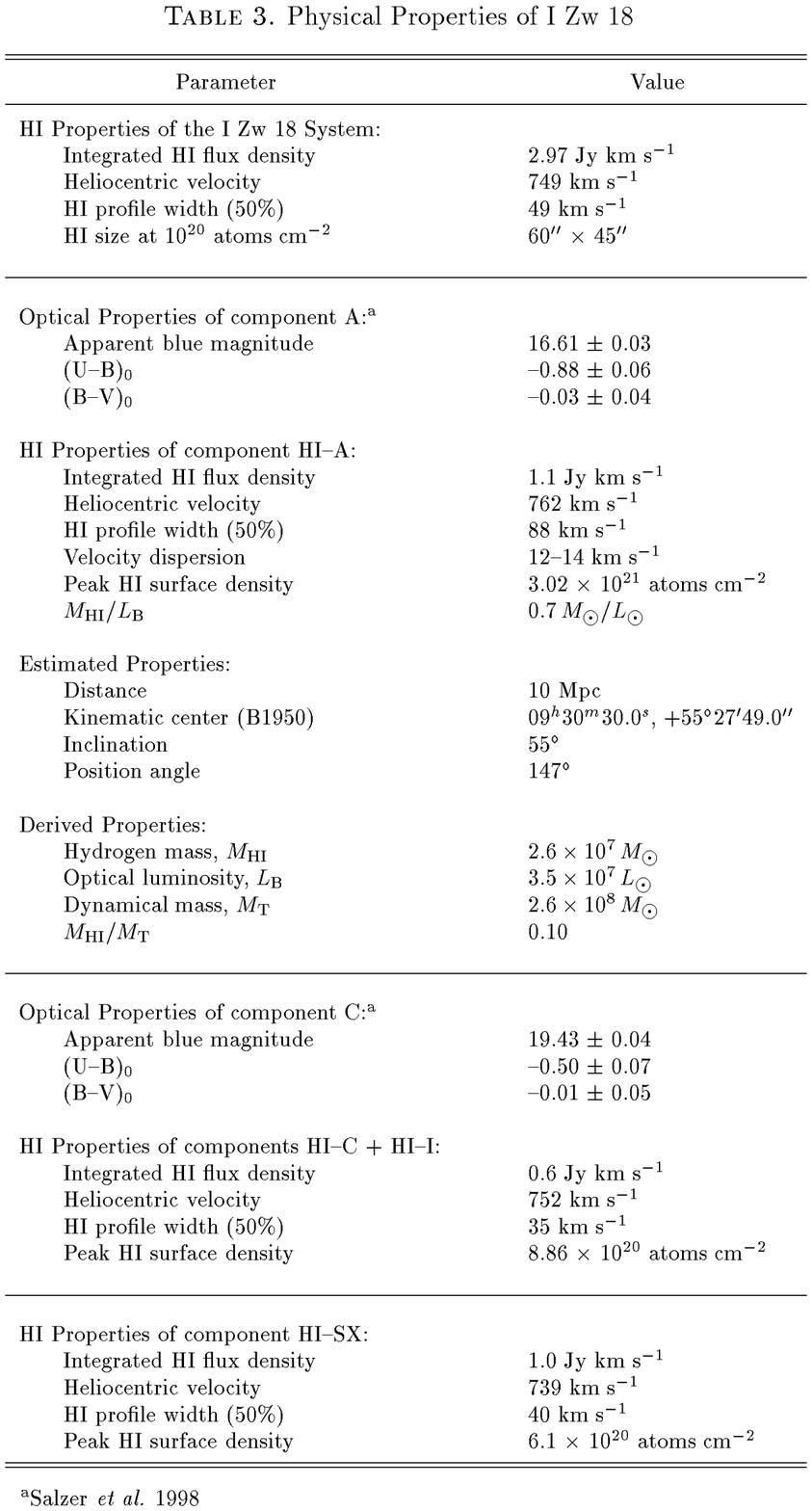,width=7.in,bbllx=50pt,bblly=1pt,bburx=600pt,bbury=715pt,clip=t}

\psfig{figure=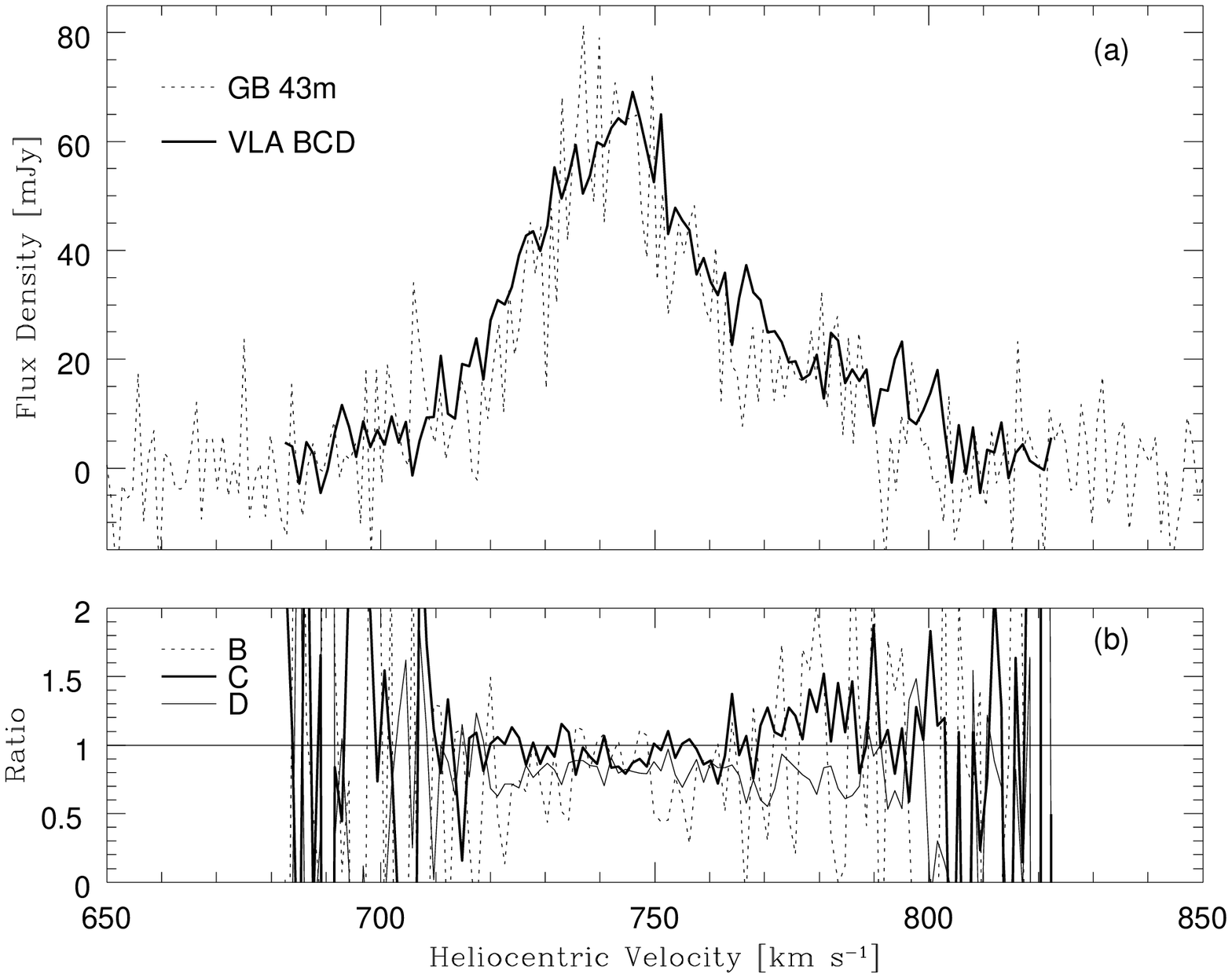,width=7.in,bbllx=1pt,bblly=150pt,bburx=600pt,bbury=700pt,clip=t}
\vskip -1. truein
\figcaption[GBflux] { (a) The global HI profile for I~Zw~18 derived from the combined VLA
data sets (see text) is denoted by the bold solid line.  The baseline subtracted HI profile
obtained from the Green Bank 43 m (Haynes \etal 1998) is denoted by the
dashed line.  Despite the absence of very short spacings,
 the VLA observations appear to have recovered all of the flux density. 
(b) The ratio between the HI profiles
for each separate configuration and the combined data.  The D configuration 
is represented by the solid line; the C configuration by the bold solid 
line; the B configuration by the dashed line.  The total flux density is recovered
in all configurations. \label{fig:gbflux} }

\psfig{figure=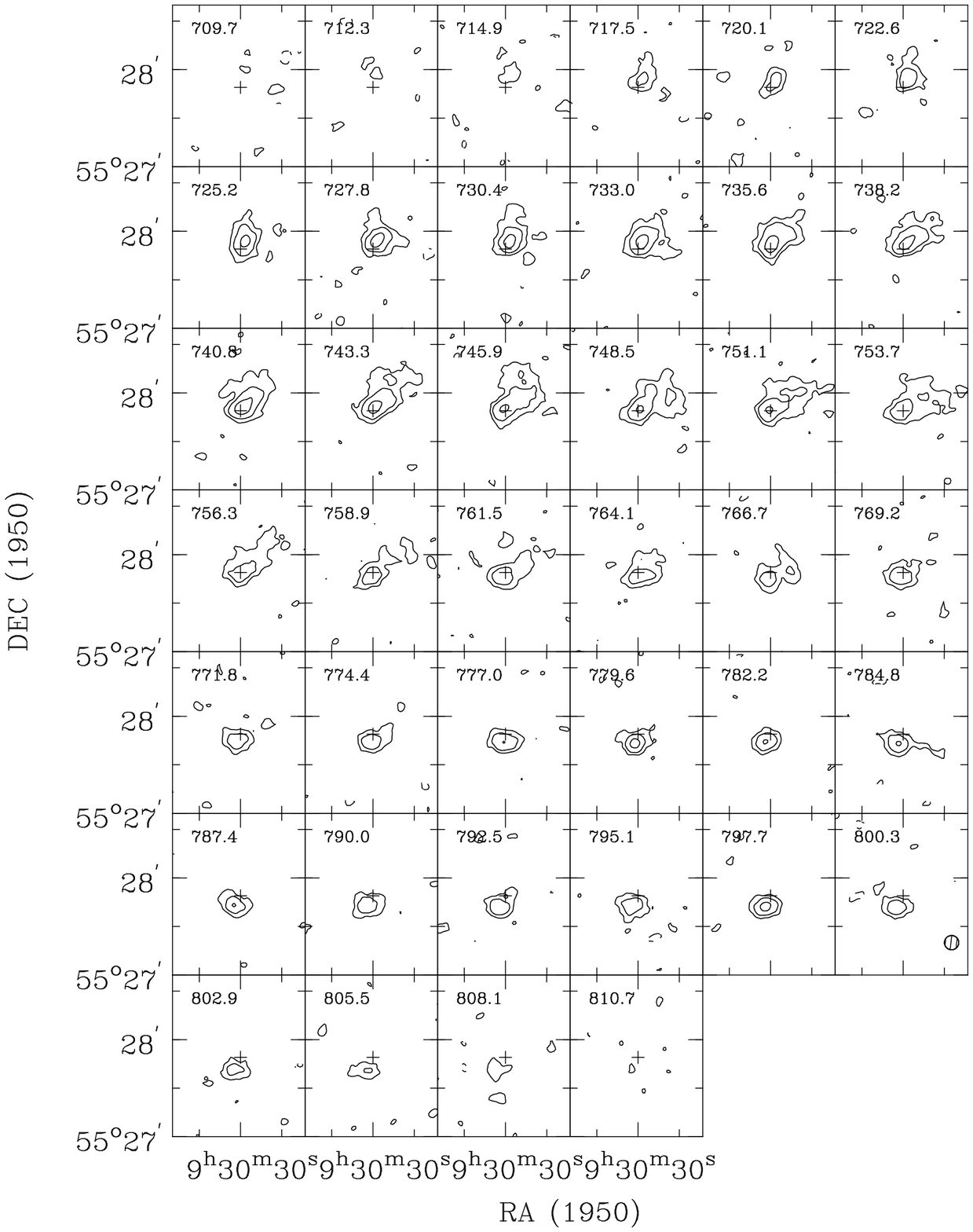,width=6.in,bbllx=70pt,bblly=20pt,bburx=600pt,bbury=700pt,clip=t}
\vskip -0.5 truein
\figcaption[chans] { Selected channels from the combined BCD data cube.   The beam
size is 8.8 $\times$ 8.3 arcsec.  The contours represent --3$\sigma$, 3$\sigma$, 
6$\sigma$, and 12$\sigma$.  The nominal kinematic center (see text) is marked with a
cross. \label{fig:chans} }

\psfig{figure=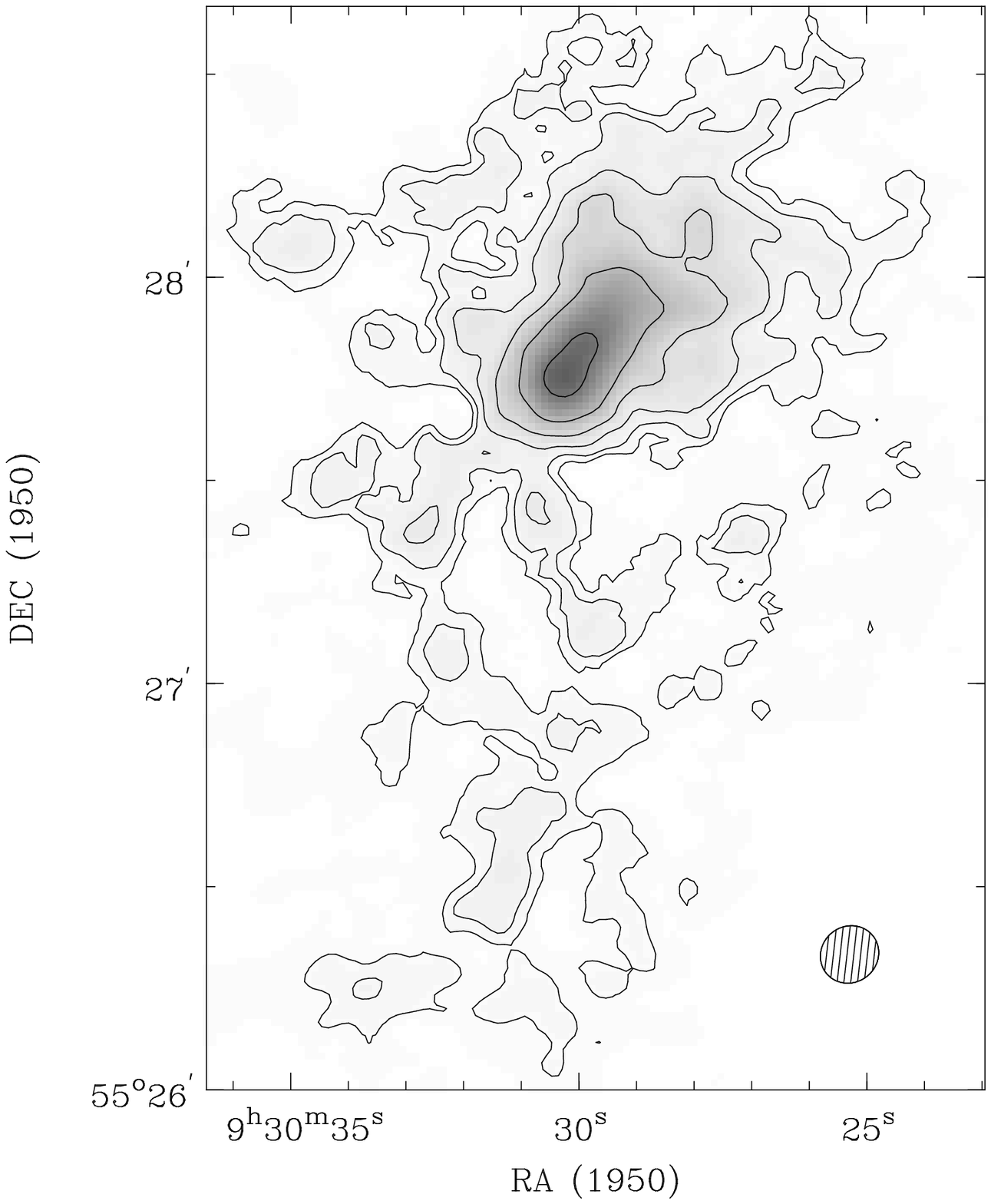,width=6.in,bbllx=1pt,bblly=10pt,bburx=600pt,bbury=700pt,clip=t}
\figcaption[Mom0] { The HI column density distribution with HI contours at
0.5, 1, 2, 4, 8, and 16 $\times$ 10$^{20}$ atoms cm$^{-2}$.  The
beam size of the HI data is 8.8 $\times$ 8.3 arcsec.  
\label{fig:mom0} }

\psfig{figure=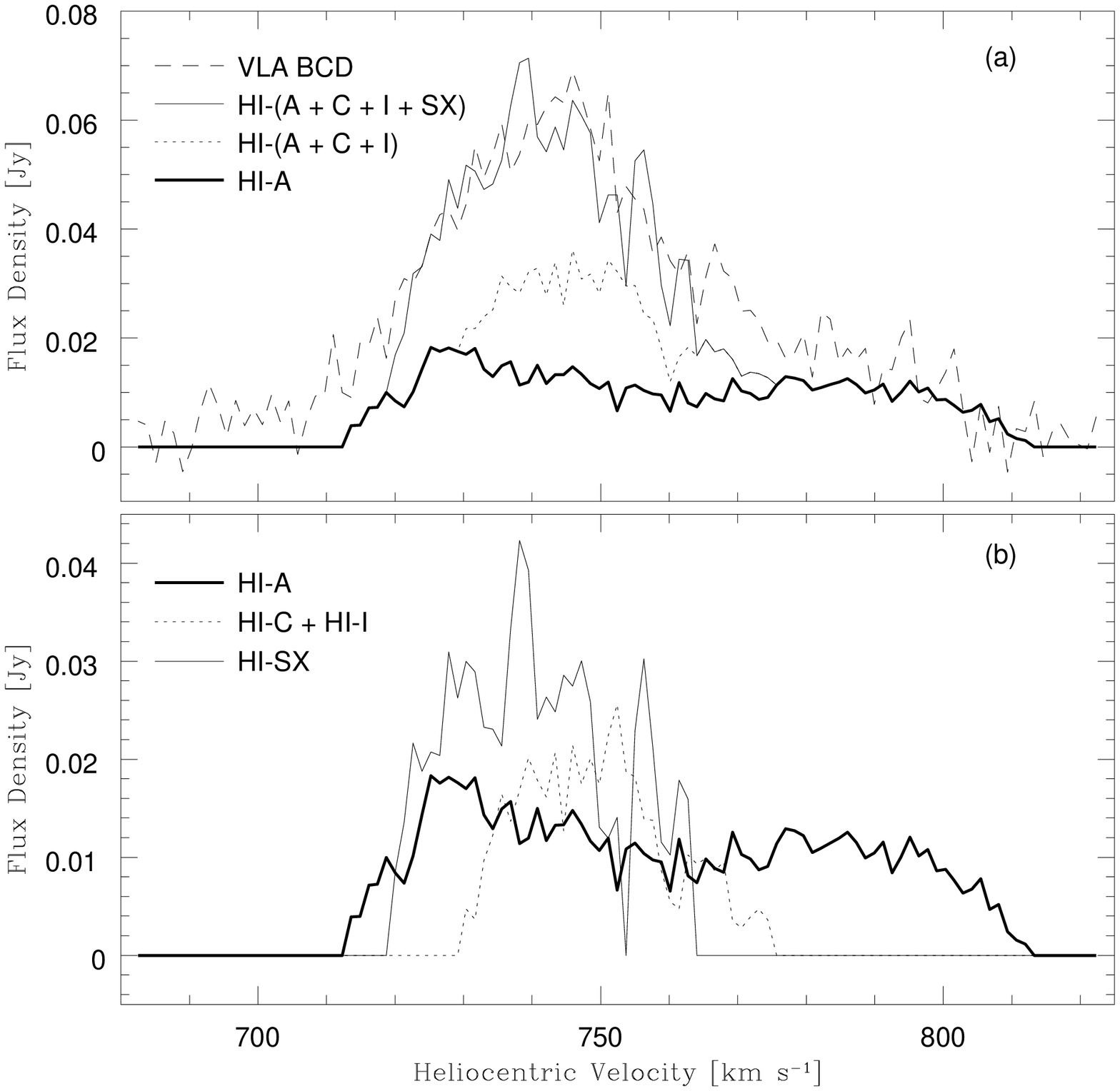,width=7.in,bbllx=1pt,bblly=150pt,bburx=600pt,bbury=700pt,clip=t}
\figcaption[flux] { Separation of HI components. (a) The asymmetric global HI profile 
derived from the combined B, C, and D configuration data cube is shown with the 
long dashed line. The flux density  associated with the main body of I~Zw~18 (HI--A) 
is denoted by the bold solid line.  The sum of the flux density associated with HI--A and the
diffuse emission to the northwest (HI--C and HI--I) is shown by the short dashed line; the sum of
HI--A, HI--C, HI--I, and the southern extension (HI--SX) is denoted by the solid line. 
(b) The HI profiles of the individual components. HI--A
has a relatively symmetric HI profile and contains approximately 40\% of the total HI flux density
while HI--SX contains $\sim$35\% of the total HI flux density.
 \label{fig:flux} }

\psfig{figure=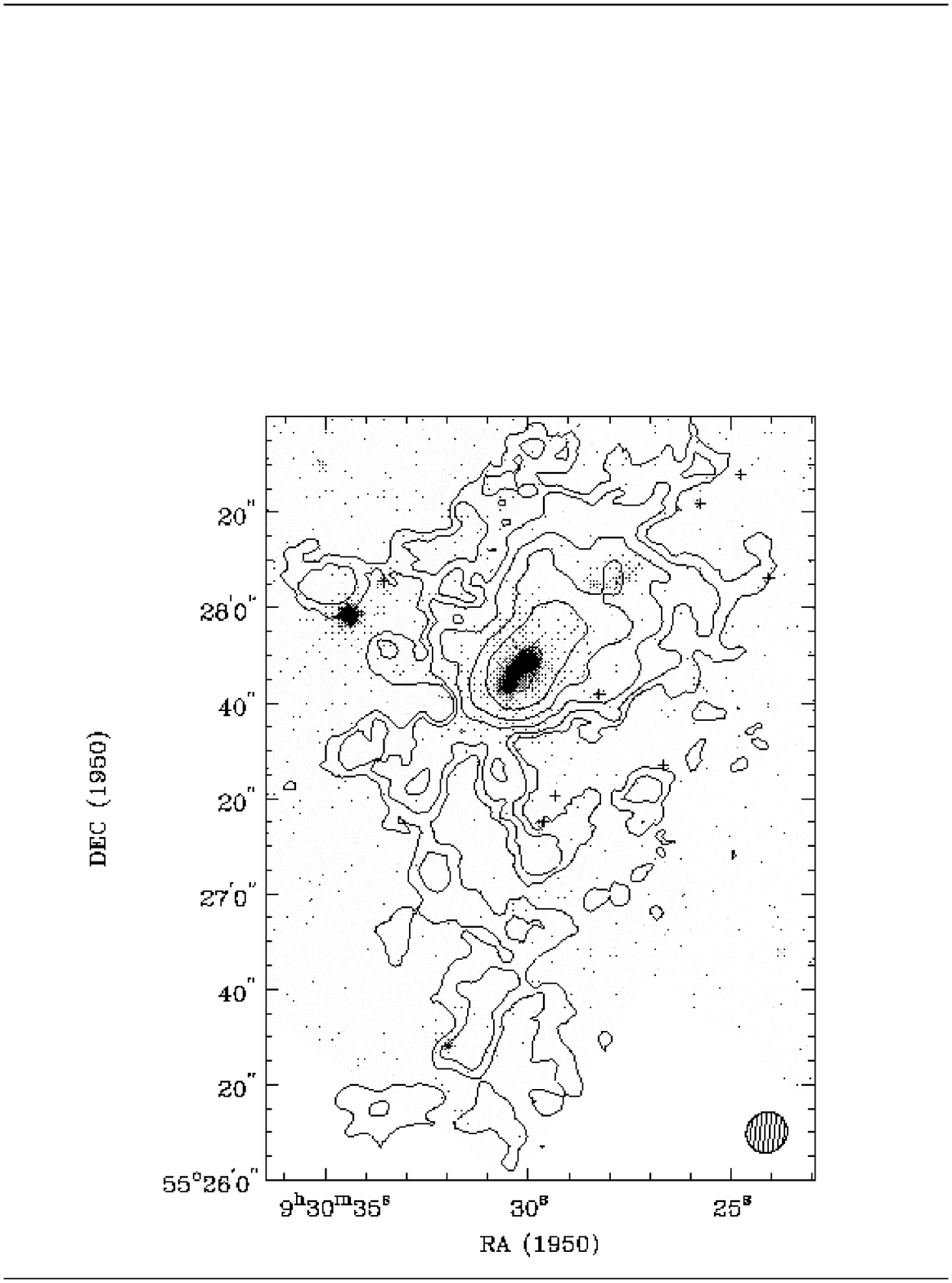,width=6.in,bbllx=1pt,bblly=-50pt,bburx=600pt,bbury=600pt,clip=t}
\figcaption[opthi] { The HI column density distribution overlayed on a mosaic of
F555W images from HST. The HI contours are 0.5, 1, 2, 4, 8, and 16 $\times$ 
10$^{20}$ atoms cm$^{-2}$.  The
beam size of the HI data is 8.8 $\times$ 8.3 arcsec.  The pixel scale of the optical
image is 0.1 arcsec/pixel.  The positions of background galaxies (used for 
registration of the HST images) are marked with crosses. Peaks in the HI column 
density are associated both with the main body of I~Zw~18 and with component C 
in the NW.  In addition, the HI distribution extends well beyond the optical systems.
\label{fig:opthi} }

\centerline{\hbox{
\psfig{figure=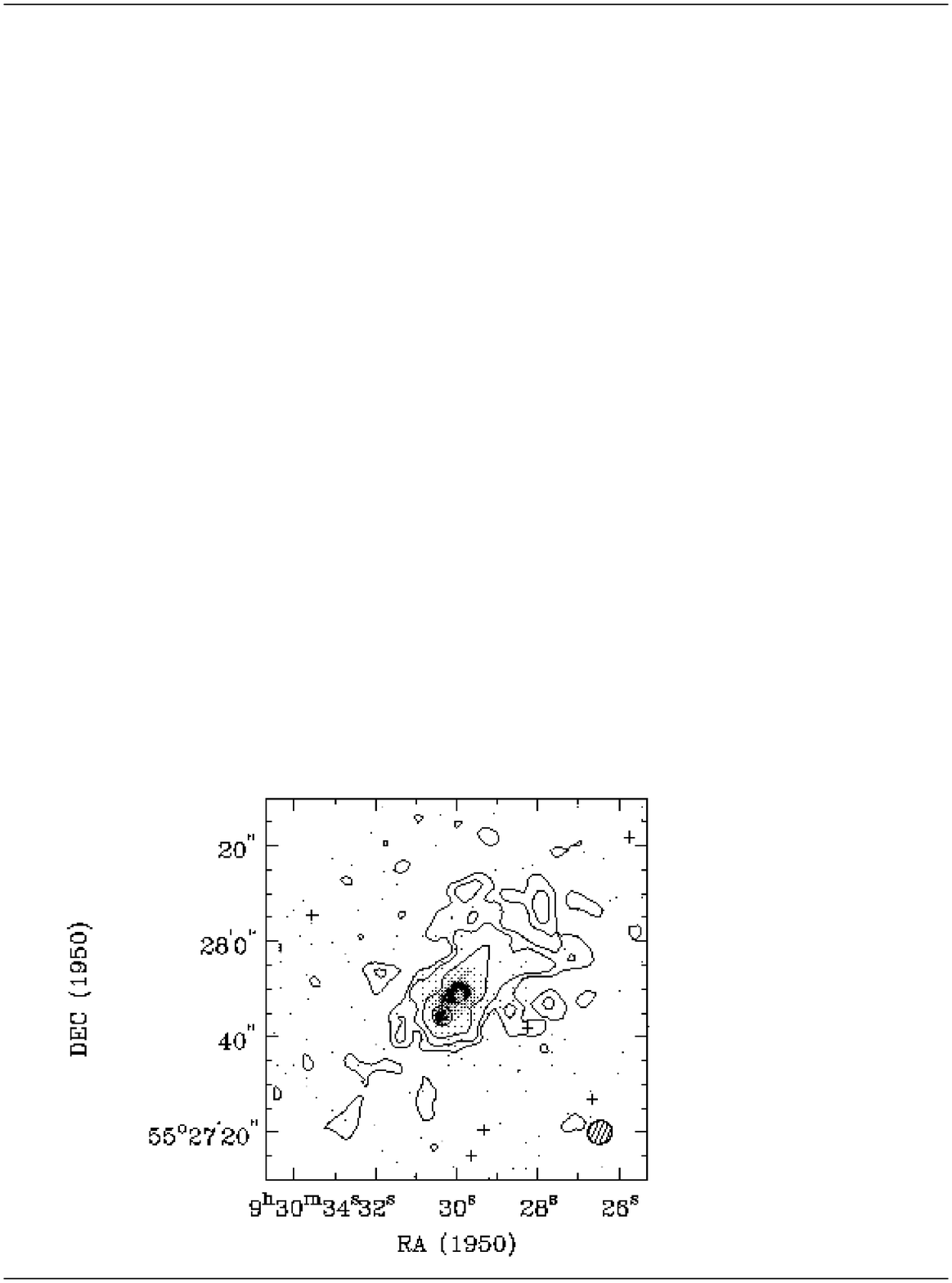,width=3.5in,bbllx=10pt,bblly=1pt,bburx=450pt,bbury=500pt,clip=t}
\psfig{figure=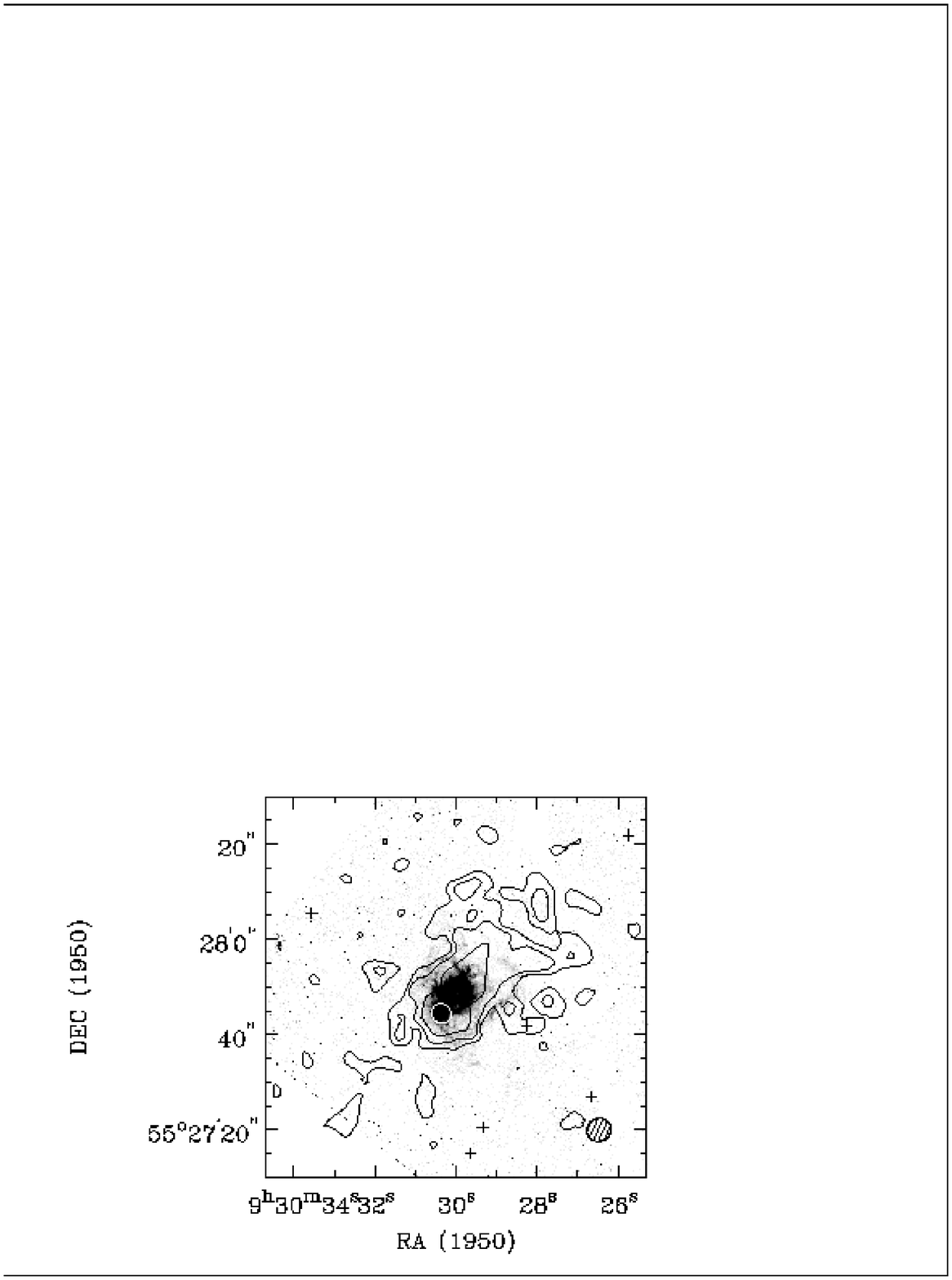,width=3.5in,bbllx=10pt,bblly=1pt,bburx=450pt,bbury=500pt,clip=t}
}}
\figcaption[hahi] {(a) The HI column density distribution overlayed on an H$\alpha$ image
(continuum subtracted F658N image from HST).
The HI contours are 3.2, 6.4, 12.8, and 25.6 $\times$ 10$^{20}$ atoms cm$^{-2}$.
The beam size of the HI data is 5.2 $\times$ 4.8 arcsec.  The
pixel scale of the optical image is 0.1 arcsec/pixel.  Peaks in the HI column density
are associated with peaks in the H$\alpha$ flux. (b) The same as (a), but with the
H$\alpha$ flux enhanced to illustrate the extent of the low surface brightness features 
in the ionized gas. \label{fig:hahi} }

\psfig{figure=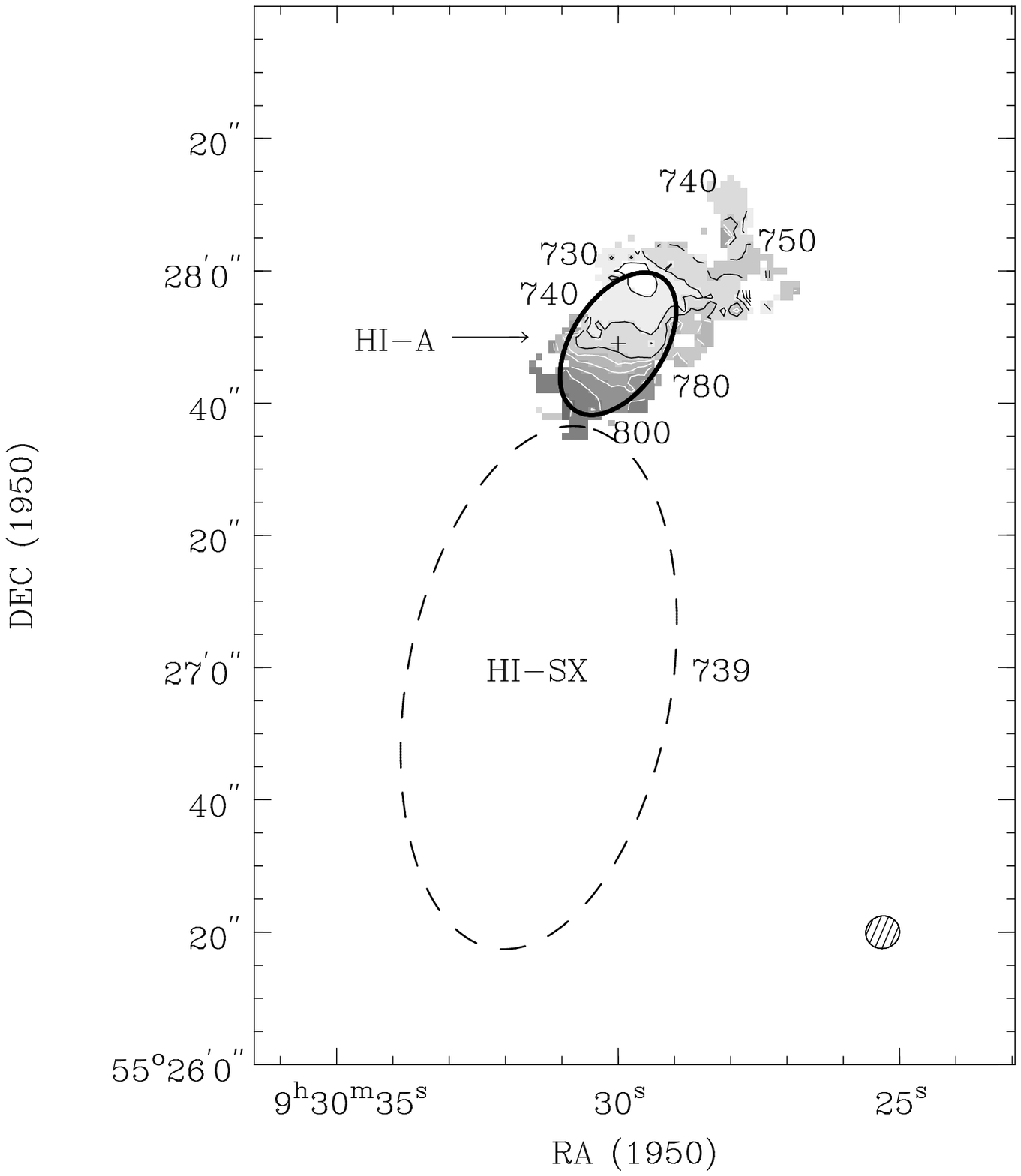,width=6.in,bbllx=1pt,bblly=10pt,bburx=600pt,bbury=700pt,clip=t}
\vskip -0.5 truein
\figcaption[velfield] {The velocity field of I~Zw~18. 
The beam size of the HI data is 5.2 $\times$ 4.8 arcsec. The contours
are separated by 10 \kms.  The southern extension is too faint to appear
in the moment map; the dashed ellipse illustrates the location of HI--SX and
its mean velocity is indicated.  HI--A is defined as the region of smoothly rotating
gas; the solid ellipse denotes the outermost ring used in the tilted--ring 
models.  HI--C appears as a kinematically distinct object to the NW of HI--A at 740 \kms; HI--I is
the ``interbody gas'' connecting HI--A and HI--C (extending west from the closed contour at
730 \kms~to HI--C). \label{fig:vel} }

\psfig{figure=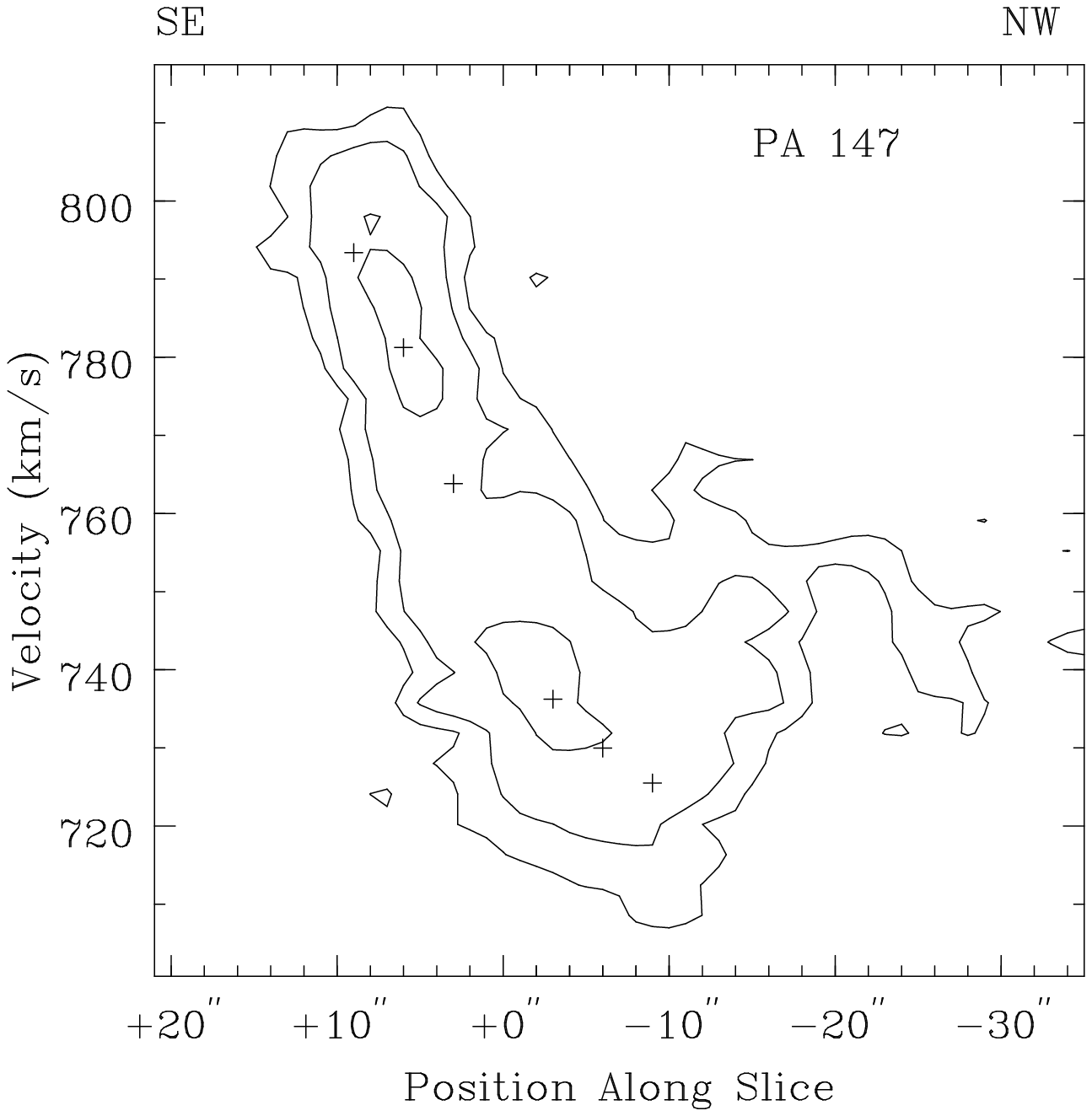,width=7.in,bbllx=1pt,bblly=10pt,bburx=600pt,bbury=600pt,clip=t}
\vskip -0.5 truein
\figcaption[pv] {Position--velocity diagram for I~Zw~18, cut at an angle of 147\arcdeg.
HI--C appears as a separate feature at 740 \kms~and 25\arcsec~to the NW.  The derived rotation
curve (based on assumed parameters) is marked with the crosses. \label{fig:pv} }

\psfig{figure=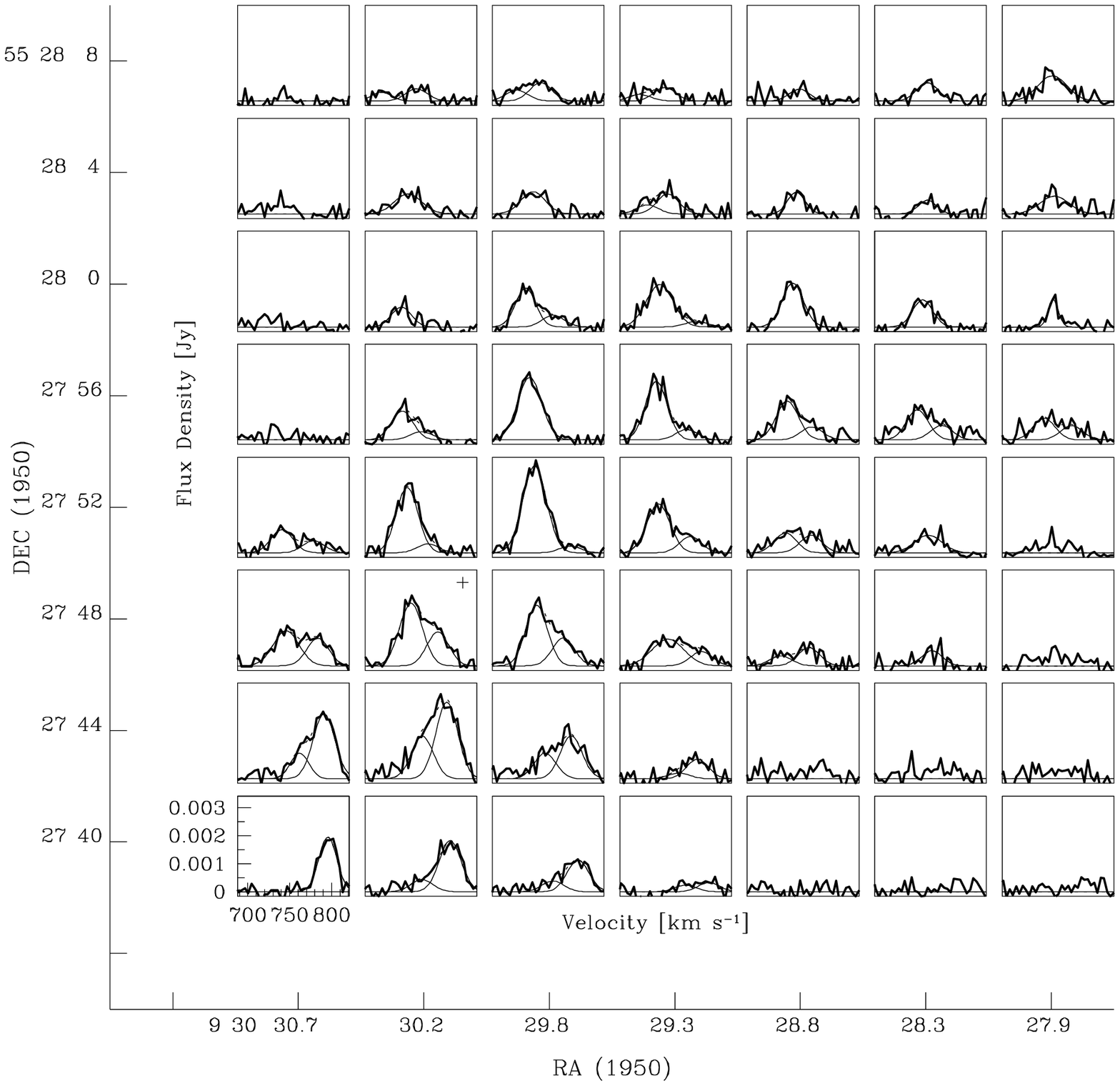,width=7.in,bbllx=1pt,bblly=150pt,bburx=600pt,bbury=700pt,clip=t}
\figcaption[profs] {Spectra from 4\arcsec~$\times$ 4\arcsec~regions of the binned
data cube.  The profiles are asymmetric, perhaps due to the overlap of the
two gas clumps associated with A(NW) and A(SE). The panel centered on the region
of the GHRS absorption spectrum is marked with a cross.  The assumed kinematic
center lies between that panel and the panel to the right. \label{fig:prfs} }

\psfig{figure=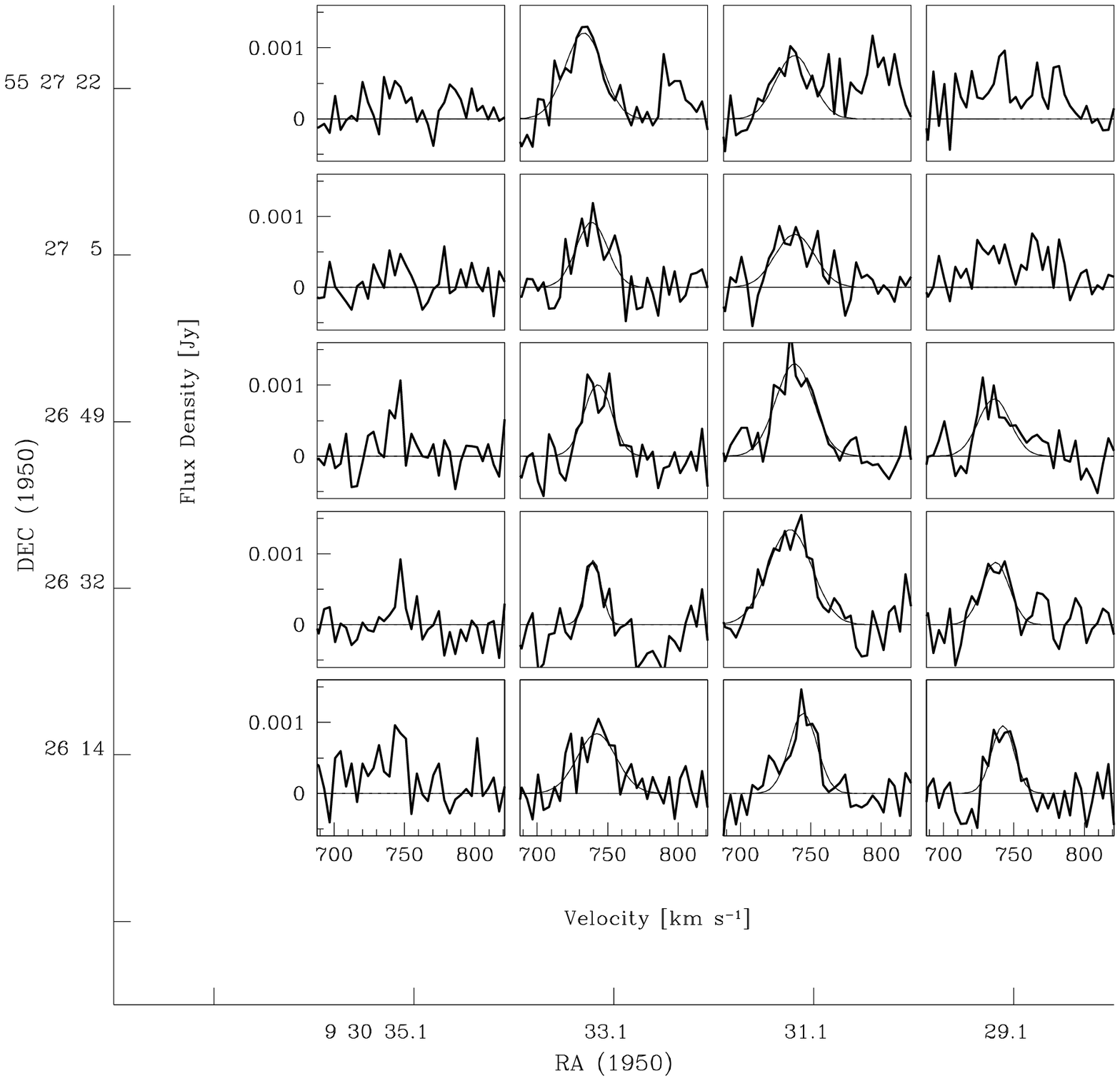,width=7.in,bbllx=1pt,bblly=150pt,bburx=600pt,bbury=700pt,clip=t}
\figcaption[southprofs] {Spectra from 17\arcsec~$\times$ 17\arcsec~regions
of the tapered, binned data cube in the region of HI--SX.  Overlayed on the profiles 
are single gaussian fits, with dispersions of about 10 \kms. HI--A begins to appear
at 800 \kms~in the center two panels of the top row.
\label{fig:sprfs} }

\psfig{figure=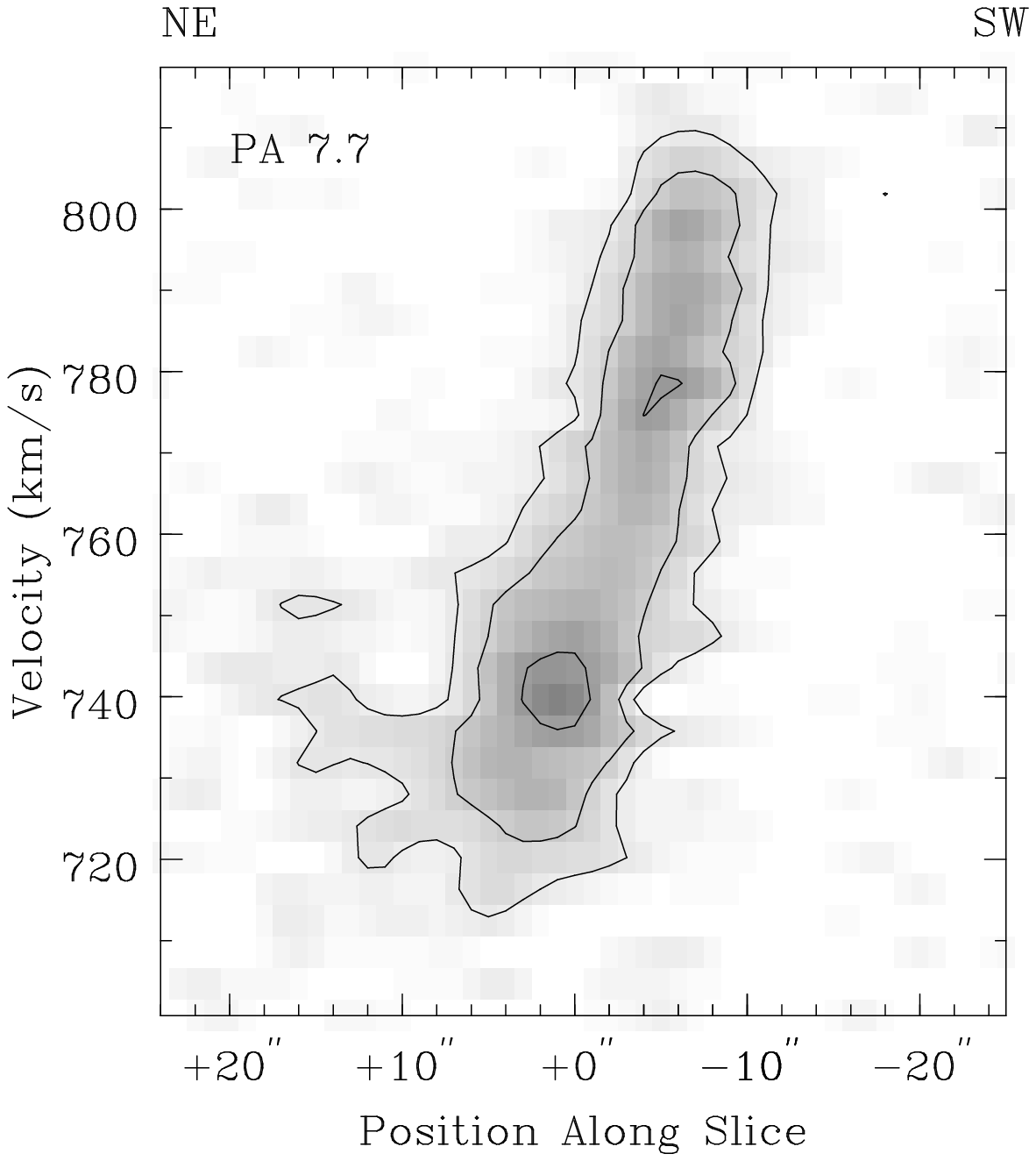,width=7.in,bbllx=1pt,bblly=10pt,bburx=600pt,bbury=600pt,clip=t}
\figcaption[pv7] {Position--velocity  diagram for I~Zw~18, cut at an angle of 7.7\arcdeg, centered at the NW HII region.  
The contours are 3$\sigma$, 6$\sigma$, 12$\sigma$ and
24$\sigma$; the lowest gray scale is at 0.5$\sigma$.
The neutral gas column density is too low in the outlying regions 
to detect features
similar to the Doppler ellipse identified by Martin (1996). \label{fig:pv7}}

\psfig{figure=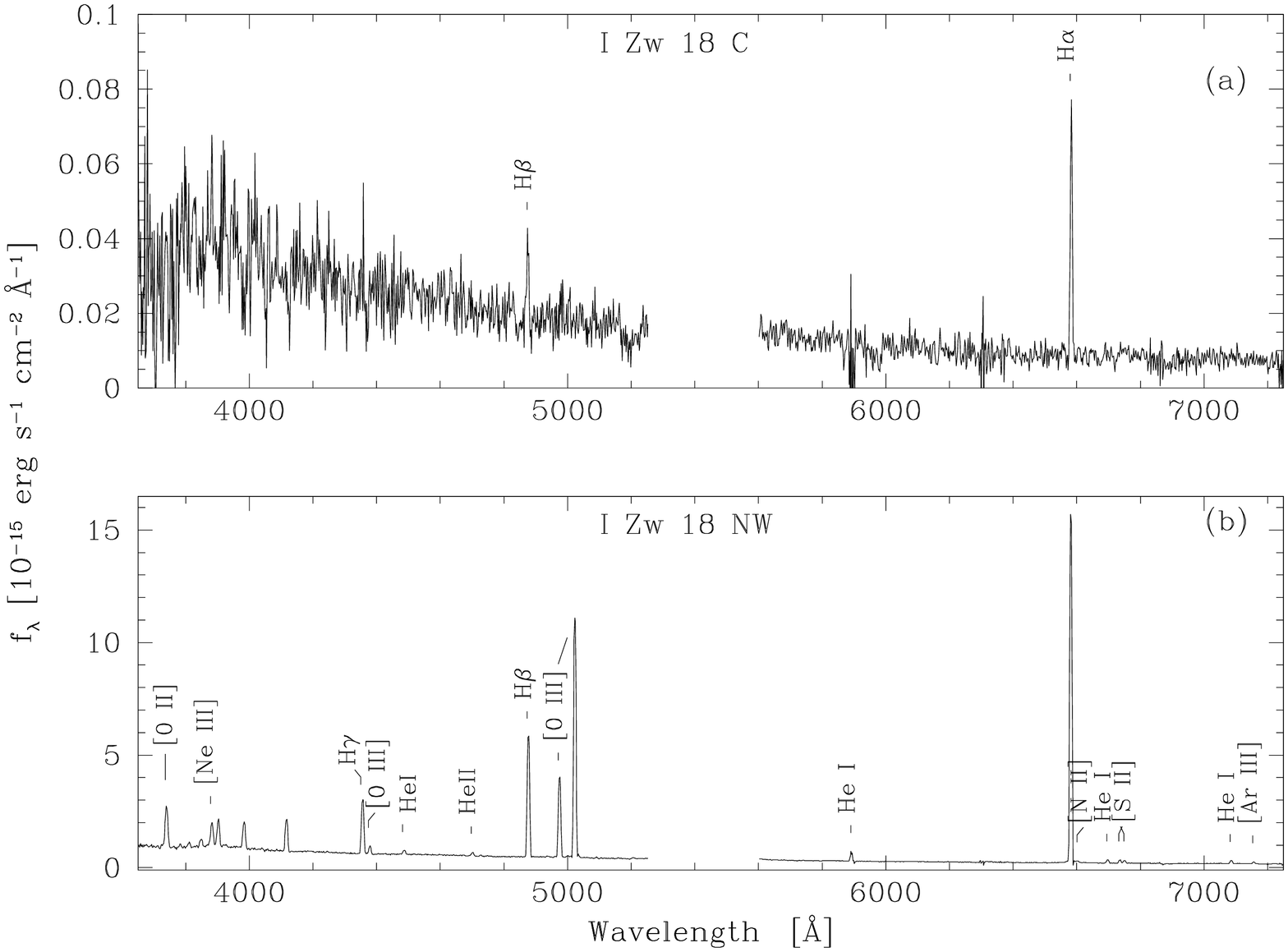,width=7.in,angle=-90.}
\figcaption[optspec] {Optical spectra of the HII region in (a) component C
and (b) the NW HII region of I~Zw~18.  \label{fig:optspec}}

\end{document}